\documentclass[12pt]{article}
\usepackage{amssymb,amsmath,epsfig,xcolor}
\usepackage{float}
\allowdisplaybreaks

\begin{document}

\title{\textbf{Astrophysical Signatures of Charged Anisotropic Compact Star
in $f(R,T)$ Gravity via Minimal Geometric Deformation}}
\author{Hira Shadab$^{1}$\thanks{hirashadab123@gmail.com}~,
M. Sharif$^{1,2}$\thanks{msharif.math@pu.edu.pk}~ and
Tayyab Naseer$^{1}$\thanks{tayyabnaseer48@yahoo.com}\\
$^1$Department of Mathematics and Statistics, The University of Lahore,\\
1-KM Defence Road Lahore-54000, Pakistan.\\
$^2$Research Center of Astrophysics and Cosmology, Khazar University, \\
Baku, AZ1096, 41 Mehseti Street, Azerbaijan.}

\date{}
\maketitle

\begin{abstract}
This work employs minimal geometric deformation to develop two new
anisotropic solutions under the effects of an electromagnetic field
in the context of $f(R,T)$ theory. Our investigation begins with a
static spherical system with an anisotropic fluid as a seed source.
After that, a new source is inserted to the existing fluid
distribution and both the fluid distributions are assumed to be
gravitationally coupled with each other. After computing the
Einstein-Maxwell field equations, a radial transformation is applied
which results in two different sets of equations. These systems are
solved individually under different constraints, generating new
solutions. To obtain the constants in our interior solutions, we
match the internal geometry with the external Reissner-Nordstr\"{o}m
spacetime at the surface. We further validate our models by using
the known mass and radius of a star LMC X-4. Our investigation
confirms that both relativistic stellar models satisfy all the
physical requirements for the specified decoupling parametric
values.
\end{abstract}

\section{Introduction}

The general theory of relativity (GR) is widely recognized as the
best model for studying gravitational problems. This approach
applies curvature and energy-momentum tensor (EMT) to construct
Einstein's field equations, which describe how matter and energy
influence spacetime geometry. Researchers have focused on the
construction of exact solutions for static spherical objects in GR
and modified theories since Schwarzschild's discovery. In 1916, he
derived the external and internal solutions for a constant density
sphere. To build a realistic picture of celestial bodies,
researchers solved gravitational field equations that satisfy
fundamental physics constraints. Obtaining exact solutions for
compact stars is challenging due to non-linear field equations, but
recent attempts have focused on constructing models for realistic
stars that correspond to different kinds of matter and geometry.

The first modification of GR is the $f(R)$ gravity theories, which
expand the geometric component of the Einstein-Hilbert action to
include a general function of the Ricci scalar. Cosmologists have
explored this framework as a potential explanation for cosmic
acceleration and early universe inflation \cite{6a}-\cite{7a}.
Analytical approaches have shown that certain $f(R)$ models are
stable under various situations \cite{8a}-\cite{9aa}. Bertolami et
al. \cite{10aa} made a significant contribution by establishing a
new link between matter and geometry. They hypothesized gravity as a
function of $R$ and $L_m$. The formulation indicated major changes
in the physical behavior of compact astrophysical objects. Some
interesting research in this field include \cite{10a, 10b}. Harko
and colleagues \cite{11aa} introduced the $f(R,T)$ theory, which
extends the Einstein-Hilbert action by defining the gravitational
Lagrangian as a function of $R$ and the trace $T$ of the EMT. The
dependence on $T$ may suggest the presence of unusual elements.
Adding $T$ disrupts energy-momentum conservation, resulting in extra
force. Test particles thus deviate from geodesic paths, resulting in
non-geodesic motion. This modified gravitational framework has
important astrophysical and cosmic implications, as proposed and
explained in \cite{15c}-\cite{33c}.

Gravitational theories use field equations to accurately explain the
self-gravitating systems. Solving these equations is challenging as
they need higher-order derivatives of geometric components.
Analytical and numerical methods provide solutions, but require
particular initial and boundary conditions. However, a comprehensive
solution requires additional information on interior physical
aspects. New factors have been identified that affect the structural
properties of compact stars \cite{r1g}-\cite{r1b}. In order to
determine the specific solution for a static anisotropic sphere,
specific metric potentials must be chosen. The system has three
independent field equations but contains five unknown values (three
matter components and two metric elements). To achieve a physically
viable outcome, limiting the geometric components to minimize
unknown factors. Researchers frequently employed the Karmarkar
condition, which establishes a relation between two metric functions
\cite{64}-\cite{67}. The vanishing of the Weyl tensor has also been
a significant topic in this context \cite{68}.

Stars are essential barrier in the history and evolution of our
galaxy. When a star's core is overheated, outward pressure is
created that overcomes gravity. A star uses balancing forces to keep
itself stable. A star will collapse if its gravitational pull is
greater than its external pressure. Black holes, neutron stars, and
white dwarfs can all result from gravitational collapse. The outcome
is largely dependent on the star's initial mass. Researchers have
studied dense formations to better understand their physical
characteristics and evolution. Neutron star is formed by the
gravitational collapse of a large star possessing a mass more than
three times that of the Sun. Their cores are dense and composed of
newly generated neutrons. Neutrons apply pressure, preventing
further gravitational collapse. Neutron stars, predicted in 1934
\cite{79}, took decades to prove due to their modest radiation
emission. Jeans' significant discovery \cite{3} demonstrated that
fluid anisotropy is governed by internal structural characteristics.
Lema\^{i}tre \cite{4} introduced the first anisotropic solution with
the constant density. Ruderman \cite{5} proposed that in dense star
interiors (more than $10^{15}~g/cm^3$), radial and tangential
pressures could differ. Anisotropy in compact stars can result from
physical factors such as pion condensation \cite{10}, extreme
magnetic fields \cite{6}-\cite{9}, nuclear interactions \cite{11pp},
and superfluid cores \cite{12}. Bowers and Liang \cite{13} proposed
that a large object's core needs to be isotropic for stability. The
study identified a significant correlation between this quantity,
the star's maximum equilibrium mass, and surface redshift.
Anisotropy has been extensively studied for its impact on stellar
properties, such as non-radial oscillations, radius, and moment of
inertia \cite{14}-\cite{38}.

New approaches, including gravitational decoupling, have been
developed to find the closed-form solution. This technique depicts
confined interiors with fluid sources like anisotropy, heat flow,
and shear stress. This approach breaks down complex field equations
from several sources into discrete sets, enabling particular
set-based solutions. The minimal geometric deformation (MGD)
methodology, developed by Jorge Ovalle \cite{41}, is a strong tool
for producing accurate stellar solutions for dense objects. It was
later expanded in \cite{42} by the same author. Casadio et al.
\cite{43} used this method to accurately calculate the Schwarzschild
spacetime. Ovalle et al. \cite{44} effectively utilized the MGD
approach to create an anisotropic version of a spherical isotropic
solution. Sharif and Naseer \cite{45b} extended the Ovalle's work
and achieved physically consistent results. Avalos and colleagues
\cite{101} proposed a concept of a static black hole with a unique
matter source. The model demonstrates how a specific element can
improve complicated features while maintaining weak energy
conditions near and beyond the event horizon. Maurya et al.
\cite{102} employed non-metricity gravity to analyze stars with
self-binding properties, employing the realistic parameters similar
to the MIT bag model. Al Habsi et al. \cite{103} utilized the MGD
technique to create anisotropic compact star solutions with minimal
complexity requirements. These investigations demonstrate how
gravitational fluids interact and exchange energy, enhancing
system's stability.

Maurya et al. \cite{104} used this approach in symmetric
teleparallel gravity and developed an anisotropic version of the
Buchdahl model. The vanishing complexity requirement was also
employed to implement this approach. Ditta et al. \cite{105}
investigated the thermodynamic properties of black holes using MGD,
including temperature, heat capacity, and free energy. This study
investigated phase transitions and entropy variations to discover
that a certain parameter and scale length, similar to a cosmological
constant, impact the stability of small black holes while
maintaining thermodynamic equilibrium. Misyura et al. \cite{106}
used MGD to Hayward and Bardeen black holes, producing non-singular
solutions with de Sitter cores and Schwarzschild-like outer regions.
The study found that anisotropic pressure can impact central
stability and proved the structures' strength against singularities
using thermodynamic analysis. Mustafa and Maurya \cite{107y}
investigated strong-field lensing in MGD-modified supermassive black
holes near galaxies like M87 and Sgr A*. The study found that
measuring angles of deflection and coefficients caused by
deformation factors minimized lensing effects compared to the
Schwarzschild model. This also gave unique signals for observation
utilizing EHT data. Other notable works in this field include
\cite{108}-\cite{125}.

This study uses the MGD scheme to extend the existing anisotropic
spherical solutions to the Einstein-Maxwell field equations to the
complex domain under $f(R,T)$ gravity scenarios. Our methodology
follows the framework given below. Section \textbf{2} covers the
principles of $f(R,T)$ modified gravity and field equations for a
static, spherically symmetric spacetime. In section \textbf{3}, the
gravitational decoupling strategy divides the field equations into
two systems, each representing a distinct fluid source. The
analytical solution for the principal fluid source is generated in
section \textbf{4}. Section \textbf{5} proposes two new solutions
for compact stars in a radially deformed system, using density and
pressure profiles as constraints. The resulting models are then
subjected to a visual analysis and their astrophysical implications
in section \textbf{6}. In section \textbf{7}, we provide our study's
key findings and show that gravitational decoupling with the MGD
procedure is useful for building realistic star models.

\section{Modified $f(R,T)$ Theory and its Fundamentals}

To describe the geometry of spacetime surrounding astronomical
objects, a metric is required. Examining how matter and energy
interact when gravity is present is crucial. Simplifying the field
equations is a major benefit of using a static spherical line
element. We thus use a spherical interior metric in this study as
\begin{equation}\label{g6}
ds^2=-A_1^2 dt^2+B_1^2 dr^2+r^2 d\Omega^2,
\end{equation}
where $A_1=A_1(r),~B_1=B_1(r)$ and $d\Omega^2=d\theta^2+\sin^2\theta
d\phi^2$.

An essential tool that elegantly explains how matter and energy
change spacetime curvature is the Einstein-Hilbert action. The
variational principle is used in classical field theory to extract
basic physical laws from an action functional. Every gravitational
theory needs an action that satisfies two conditions: \emph{(i)} it
must be constant across coordinate systems, and \emph{(ii)} it must
be based on the Ricci scalar, which is the most basic scalar
quantity that captures spacetime curvature. In $f(R,T)$ theory, the
action function is expressed as
\begin{equation}\label{g2a}
I=\int\sqrt{-g}\left(\frac{f(R,T)}{16\pi}+L_{m}+L_{E}+\omega
L_{\Psi}\right)d^{4}x,
\end{equation}
where $L_m$ and $L_E$ denote the Lagrangian densities associated
with the electromagnetic field and matter distribution,
respectively. We add a new source term, represented by $L_{\Psi}$,
that is gravitationally connected to the initial matter
distribution. The decoupling parameter $\omega$ quantifies the
effect of a new source on the physical characteristics of
self-gravitating structure. We obtain the following system of field
equations by applying the variational principle to Eq.\eqref{g2a} as
\begin{equation}\label{g2}
G_{\alpha\eta}=8\pi T_{\alpha\eta}^{(eff)},
\end{equation}
with
\begin{equation}\label{z1}
T_{\alpha\eta}^{(eff)}=\bar{T}_{\alpha\eta}+\omega
\Psi_{\alpha\eta},
\end{equation}
where the Einstein tensor,
$G_{\alpha\eta}=R_{\alpha\eta}-\frac{1}{2}R g_{\alpha\eta}$,
represents the core curvature of spacetime. Additionally, the Ricci
curvature tensor and metric tensor are represented by
$R_{\alpha\eta}$ and $g_{\alpha\eta}$, respectively. The two
distinct sources are connected by the constant $\omega$ and
$\Psi_{\alpha\eta}$ describing an additional gravitational source.
The modified EMT is represented by the expression
$\bar{T}_{\alpha\eta}$ and can be further categorized as
\begin{equation}\label{1ab}
\bar{T}_{\alpha\eta}=\frac{1}{f_{R}}\big(T_{\alpha\eta}+E_{\alpha\eta}\big)+T_{\alpha\eta}^{(D)},
\end{equation}
where $T_{\alpha\eta}$ corresponds to the standard matter EMT,
$E_{\alpha\eta}$ represents the electromagnetic component of the
EMT, and $T_{\alpha\eta}^{(D)}$ reflects the extra contributions
resulting from changes to the action function. The mathematical
expression of the correcting terms is defined as
\begin{eqnarray}
\nonumber T _{\alpha\eta}^{(D)}&=&\frac{1}{8\pi f_{R}}\bigg[f_{T}
T_{\alpha\eta}+\bigg\{\frac{R}{2} \bigg(\frac{f}{R}-f_{R}\bigg)-
L_{m}f_{T}\bigg\}g_{\alpha\eta}\\\label{2a}
&-&(g_{\alpha\eta}\Box-\nabla_{\alpha}\nabla_{\eta})f_{R}+2f_{T}g^{\sigma\nu}\frac{\partial^2
L_{m}}{\partial g^{\alpha\eta}\partial g^{\sigma\nu}}\bigg].
\end{eqnarray}

Researchers have extensively studied anisotropic fluid interiors in
numerous physical systems. Anisotropic fluids differ from isotropic
matter by exhibiting diverse physical properties in different
directions. This distinction is critical for studying massive
phenomena, such as relativistic stars or areas with high gravity.
Our research focuses on anisotropic compact structures, hence the
EMT for their interior material composition is expressed as
\begin{equation}\label{g5}
T_{\alpha\eta}=(p_t+\rho)u_\alpha u_\eta-p_t
g_{\alpha\eta}-(p_t-p_r)v_\alpha v_\eta.
\end{equation}
The variables $\rho,~p_r,~p_t,~u_\alpha$, and $v_\alpha$ correspond
to energy density, radial pressure, tangential pressure,
four-velocity, and four-vector, respectively. In a co-moving
reference frame, the metric \eqref{g6} simplifies the final two
physical quantities as follows
\begin{equation}\nonumber
u_\alpha=(-A_1,0,0,0), \quad v_\alpha=(0,B_1,0,0),
\end{equation}
assuring satisfaction of the conditions listed below
$$u^\alpha u_{\alpha}=-1, \quad v^\alpha
v_{\alpha}=1, \quad u^\alpha v_{\alpha}=0.$$

The mathematical expression for the electromagnetic EMT can be
represented as follows
\begin{equation}\label{aaa}
E_{\alpha\eta}=-\frac{1}{4\pi}\bigg[\frac{1}{4}g_{\alpha\eta}F^{\nu\mu}
F_{\nu\mu}-F^{\nu}_{\alpha} F_{\eta\nu} \bigg],
\end{equation}
whereas the Maxwell field tensor is expressed by
$F_{\eta\nu}=\Phi_{\nu;\eta}-\Phi_{\eta;\nu}$ with
$\Phi_\eta=\Phi(r)\delta^0_\eta$ representing the four-potential.
The Maxwell equations can be represented in a compact form as
\begin{equation}\label{aa}
F^{\alpha\eta}_{~~;\eta}=4\pi \Theta^\alpha, \quad
F_{[\alpha\eta;\nu]}=0.
\end{equation}
The formula $\Theta^\eta=\jmath u^\eta$ denotes current density,
while $\jmath$ represents charge density. Using the metric
\eqref{g6}, we may obtain the following differential equation from
Eq.\eqref{aa} as
\begin{equation}\label{aabb}
\Phi^{\prime\prime}-\frac{1}{2r}\bigg\{{r \bigg(\frac
{A_1^{\prime}}{A_1}+\frac
{B_1^{\prime}}{B_1}\bigg)-4}\bigg\}\Phi^\prime=4\pi\jmath A_1B_1^2.
\end{equation}
Here, prime denotes the derivative of a radial coordinate $r$.
Integrating Eq.\eqref{aabb} yields a first-order differential
equation as
\begin{equation}
\Phi^\prime= \frac{q A_1B_1}{r^2}, \quad q=\int^r_0\jmath A_1B_1
\bar{r}^2d\bar{r},
\end{equation}
where $'=\frac{\partial}{\partial r}$ and $q \equiv q(r)$ denotes
the overall charge within the interior. The Maxwell equations in
classical electromagnetism state that a stationary charge generates
an electric field and moving charges create a magnetic field. The
interaction between electricity and magnetism happens when charges
move. Even with a static system, electromagnetic equations
\eqref{aaa} require both electric and magnetic fields to completely
describe the system. Our study found that the combination of these
two components produces electromagnetic effects. The magnetic field
is implied in our calculations due to the movement of charges under
gravity $f(R,T)$.

Selecting a suitable model for extended theory of gravity is crucial
for achieving meaningful results in astrophysics and cosmology. This
technique enables researchers to describe a wide range of
astrophysical occurrences beyond GR. Acceptance of the chosen has a
significant impact on developing new experimental methods in
astrophysics and understanding present observational data. The
linear version of the $f(R,T)$ gravity is represented as
\begin{equation}\label{3ac}
f(R,T)=f_1(R)+ f_2(T)=R+2\zeta T,
\end{equation}
with an arbitrary parameter $\zeta$. This technique is particularly
effective for examining the issues related to the cosmological
constant $\Lambda$. Recent advancements in running vacuum models and
scale-dependent gravitational frameworks indicate that an effective
cosmic parameter could address issues related to $\Lambda$
\cite{105a}-\cite{105e}. The link between the Hubble parameter and
$\Lambda$ is denoted as $\Lambda_{eff}\propto H^{2}$ \cite{11}. The
physical interpretation and effects of $T$ can lead to important
theoretical improvements, leading to new insights into astrophysical
systems and cosmic structure \cite{107}. Using the metric \eqref{g6}
and fluid description \eqref{g5}, the independent components of the
field equations \eqref{g2} are calculated as
\begin{align}\label{g8}
8\pi(\rho-\omega \Psi^{0}_{0})+\zeta \bigg(3
\rho-\frac{p_r}{3}-\frac{2 p_t}{3}
\bigg)+\frac{q^2}{r^4}&=-\frac{1}{r^2B_1^2}+\frac{2B_1'}{rB_1^3}+\frac{1}{r^2},
\\\label{g9}8\pi({p}_r+\omega\Psi^{1}_{1})- \zeta
\bigg(\rho-\frac{7 p_r}{3}-\frac{2 p_t}{3}
\bigg)-\frac{q^2}{r^4}&=\frac{2A_1'}{rA_1 B_1^2}+\frac{1}{r^2
B_1^2}-\frac{1}{r^2},
\\\nonumber 8\pi({p}_t+\omega \Psi^{2}_{2})-\zeta
\bigg(\rho-\frac{p_r}{3}-\frac{8 p_t}{3}
\bigg)+\frac{q^2}{r^4}&=-\frac{B_1'}{r B_1^3}+\frac{1}{A_1
B_1^2}\\\label{g10} &\times\bigg(A_1''-\frac{A_1'
B_1'}{B_1}+\frac{A_1'}{r}\bigg).
\end{align}

The mass function indicates the amount of matter within a particular
distance from the center of a spherically symmetric spacetime. This
is especially essential when investigating dense stars. The
Misner-Sharp mass in the form of the electric charge is calculated
as \cite{faj}
\begin{align}\nonumber
m(r)&=4\pi\int_0^r r^2\rho(r)dr+\int_0^r\frac{qq'}{{r}}d{r}
\\\label{g12a}&=4\pi\int_0^r\rho{r}^2d{r}+\frac{1}{2}\int_0^r\frac{q^2}
{{r}^2}d{r}+\frac{q^2}{2r}.
\end{align}
The complexity of the energy density of the stellar model determines
the method (analytical or numerical) for solving the above
differential equation.

\section{Minimal Gravitational Decoupling}

When an extra source is introduced to the initial anisotropic fluid,
the related field equations become more complex and challenging to
solve. The number of unknowns increases to
$(A_1,B_1,\rho,p_r,p_t,\Psi_{0}^{0},\Psi_{1}^{1},\Psi_{2}^{2})$. To
achieve a precise solution, the degrees of freedom must be reduced
by appropriate approaches or physical limitations. To get a unique
solution, we apply gravitational decoupling \cite{33aa} to the field
equations. This approach simplifies equations and creates a new
reference frame for temporal and radial metric potentials. As a
result, we consider the following line element as a solution to
Eqs.\eqref{g8}-\eqref{g10}, expressed as
\begin{equation}\label{g15}
ds^2=-A_2^2(r) dt^2+\frac{1}{B_2^2(r)}dr^2+r^2d\Omega^2.
\end{equation}
The metric components undergo a linear transformation under the
considered technique, which is expressed as
\begin{equation}\label{g16}
A_2\rightarrow A_1= A_2+\omega s, \quad B_2\rightarrow B_1=
B_2+\omega t.
\end{equation}
The radial $t$ and temporal deformation function $s$ are included in
the above expressions, and changing the decoupling parameter
$\omega$ to regulate the deformation. Our chosen MGD method produces
$t \rightarrow \bar t$ and $s \rightarrow 0$ by just deforming the
$g_{rr}$ component and leaving the $g_{tt}$ potential unchanged.
Consequently, Eq.\eqref{g16} turns into
\begin{equation}\label{g17}
A_2\rightarrow A_1= A_2, \quad B_2\rightarrow B_1=
B_2+\omega\bar{t},
\end{equation}
where $\bar{t}=\bar{t}(r)$. It is crucial to understand that
spherical symmetry is preserved by this linear translation. Two
distinct systems are produced by applying the mapping \eqref{g17} to
Eqs.\eqref{g8}-\eqref{g10}. The seed fluid at $\omega=0$ is
represented by the first system, which is obtained as follows
\begin{align}\label{g8aa}
8\pi\rho+\zeta \bigg(3 \rho-\frac{p_r}{3}-\frac{2 p_t}{3}
\bigg)+\frac{q^2}{r^4}&=-\frac{1}{r^2B_2^2}+\frac{2B_2'}{rB_2^3}+\frac{1}{r^2},
\\\label{g9aa}8\pi{p}_r- \zeta
\bigg(\rho-\frac{7 p_r}{3}-\frac{2 p_t}{3}
\bigg)-\frac{q^2}{r^4}&=\frac{2A_2'}{rA_2 B_2^2}+\frac{1}{r^2
B_2^2}-\frac{1}{r^2},
\\\label{g10aa}8\pi{p}_t-\zeta
\bigg(\rho-\frac{p_r}{3}-\frac{8 p_t}{3}
\bigg)+\frac{q^2}{r^4}&=-\frac{A_2'B_2'}{A_2 B_2^3}+\frac{A_2'}{rA_2
B_2^2}+\frac{A_2''}{A_2 B_2^2}-\frac{B_2'}{rB_2^3},
\end{align}
from which we explicitly extract the following fluid variables as
\begin{align}\nonumber
\rho&=\frac{-1}{12 (\zeta +2 \pi ) (\zeta +4 \pi ) r^4 A_2
B_2^3}\big[\zeta r^4 A_2' B_2'-\zeta r^4 B_2 A_2''-2 \zeta r^3 B_2
A_2'\\\nonumber &-8 \zeta r^3 A_2 B_2'-24 \pi r^3 A_2 B_2'+5 \zeta
q^2 A_2 B_2^3+12 \pi q^2 A_2 B_2^3-4 \zeta r^2 A_2
B_2^3\\\label{g18} &+4 \zeta r^2 A_2 B_2-12 \pi r^2 A_2 B_2^3+12 \pi
r^2 A_2 B_2\big],\\\nonumber p_r&=\frac{-1}{12 (\zeta +2 \pi )
(\zeta +4 \pi ) r^4 A_2 B_2^3}\big[\zeta r^4 B_2 A_2''-\zeta r^4
A_2'B_2'-10 \zeta  r^3 B_2 A_2'\\\nonumber &-24 \pi r^3 B_2 A_2'-4
\zeta r^3 A_2 B_2'-5 \zeta q^2 A_2B_2^3-12 \pi q^2 A_2 B_2^3+4 \zeta
r^2 A_2 B_2^3\\\label{g19} &-4 \zeta r^2 A_2 B_2+12 \pi r^2 A_2
B_2^3-12 \pi r^2 A_2 B_2\big],\\\nonumber p_t&=\frac{-1}{12 (\zeta+2
\pi) (\zeta +4 \pi ) r^4 A_2 B_2^3}\big[5 \zeta r^4 A_2' B_2'-5
\zeta r^4 B_2A_2''-12 \pi r^4 B_2 A_2''\\\nonumber &+12 \pi r^4 A_2'
B_2'-4 \zeta r^3 B_2 A_2'-12 \pi r^3 B_2 A_2'+2 \zeta r^3 A_2
B_2'+12 \pi r^3 A_2 B_2'\\\label{g20} &+7 \zeta q^2 A_2 B_2^3+12 \pi
q^2 A_2 B_2^3-2 \zeta r^2 A_2 B_2^3+2 \zeta r^2 A_2 B_2\big].
\end{align}

Additionally, the influence of the newly added source term
$\Psi_{\alpha\eta}$ when $\omega=1$ is given by the following set of
equations (obtained by using the binomial theorem up to first order
of the parameter $\omega$, i.e., considering just $O(\omega)$ terms)
as
\begin{align}\label{g21}
8\pi \Psi_{0}^{0}&=\frac{2\bar{t}}{r B_2^3}\bigg(\frac
{3B_2'}{B_2}-\frac{1}{r}\bigg)-\frac{2\bar t'}{r
B_2^3},\\\label{g22} 8\pi \Psi_{1}^{1}&=\frac{2\bar{t}}{r
B_2^3}\bigg(-\frac {2A_2'}{A_2}-\frac{1}{r}\bigg),\\\label{g23} 8\pi
\Psi_{2}^{2}&=-\frac{\bar t'}{B_2^3}\bigg(\frac
{A_2'}{A_2}+\frac{1}{r}\bigg)+\frac{\bar{t}}{B_2^3}\bigg(\frac
{3A_2' B_2'}{A_2 B_2}-\frac{2A_2'}{r
A_2}-\frac{2A_2''}{A_2}+\frac{3B_2'}{r B_2}\bigg).
\end{align}
Five unknown quantities ($\rho, p_r, p_t, A_2, B_2$) exist in the
system \eqref{g18}-\eqref{g20}. Two conditions must be chosen
randomly in order to produce a suitable analytical solution. Four
variables ($\bar{t},\Psi_{0}^{0},\Psi_{1}^{1},\Psi_{2}^{2}$) are
unknown in the second set \eqref{g21}-\eqref{g23}. The sector
connected to the $\Psi$ functions will be subjected to a unique
constraint in order to solve this. At this point, the physical
factors are described as
\begin{align}\label{g13}
\bar{\rho}=\rho-\omega \Psi_{0}^{0},\quad \bar{p}_{r}=p_r+ \omega
\Psi_{1}^{1},\quad \bar{p}_{t}=p_t+\omega \Psi_{2}^{2},
\end{align}
along with the resulting anisotropy produced by
\begin{equation}\label{g14}
\bar{\Delta}=\bar{p}_{t}-\bar{p}_{r}=(p_t-p_r)
+\omega(\Psi_{2}^{2}-\Psi_{1}^{1})=\Delta+\Delta_{\Psi},
\end{equation}
which, either positive or negative, impacts the stability of a
compact star. Positive anisotropy, where tangential pressure exceeds
radial pressure, promotes stability by creating an outward force.
The star's outward repulsive force counteracts gravity and prevents
further collapse. Negative anisotropy, or tangential pressure lower
than radial pressure, can produce instability. This stage decreases
outward pressure, making the star more susceptible to gravitational
collapse. Stars with negative anisotropy have a lower stability
range than stars with positive anisotropy.

\section{Closed-form Solution of Seed Fluid Content}

Einstein's field equations have been studied theoretically,
resulting in analytical and numerical solutions. The study focuses
two sets of field equations. Solving the first set of non-linear
differential equations requires balancing the number of equations
and unknown variables under two unique restrictions. To do this, we
apply a well-known radial metric component firstly, as shown below
\cite{88aa}
\begin{equation}\label{g52}
B_2^2=\bigg(1-\frac{r^2}{d^2}\bigg)^{-4}.
\end{equation}
To ensure dimensional consistency, the constant $d^2$ should have
the dimension of length squared. The system's anisotropy can be
represented using Eqs.\eqref{g9} and \eqref{g10} as
\begin{align}\nonumber
\Delta(r)&=8\pi(p_t-p_r)=-\frac{4 \pi }{(\zeta +4 \pi) r^4 A_2
B_2^3} \big[r^3 \big\{B_2\big(A_2'-r A_2''\big)
\\\label{g12d} &+r A_2' B_2'\big\}+A_2 \big\{r^3 B_2'+B_2^3 \big(2
q^2-r^2\big)+r^2 B_2\big\}\big],
\end{align}
which has zero value at the center. Using Eqs.\eqref{g52} and
\eqref{g12d} yields
\begin{align}\nonumber
\Delta&=\frac{1}{2 d^8 (\zeta +4 \pi ) r^4 A_2}[A_2(-2 d^8 q^2+6 d^4
r^6-8 d^2 r^8+3 r^{10})\\\label{10} &-r^3 (d^2-r^2)^3\{(d^2+3 r^2)
A_2'+r(r^2-d^2) A_2''\}].
\end{align}
After rearranging Eq.\eqref{10}, it becomes
\begin{align}\nonumber
&\frac{\{r^3(d^2-r^2)^3(d^2+3r^2)\}
A_2'}{\{r^4(d^2-r^2)^3(r^2-d^2)\}A_2}
+\frac{A_2''}{A_2}=\frac{-1}{r^4\big(d^2-r^2\big)^3\big(r^2-d^2\big)}\\\label{g53i}&\times[2d^8
r^4(4\pi+\zeta)\Delta+2d^8q^2-6d^4r^6+8 d^2r^8-3r^{10}].
\end{align}

Solving Eq.\eqref{g53i} produces the metric function $A_2$ for a
particular anisotropy. For a physically valid solution, anisotropy
must be zero at the core and positive elsewhere. We choose an
anisotropy that fits all of the requirements given above as
\begin{equation}\label{g53c}
\Delta=\frac{-2d^8q^2+6d^4r^6-8d^2r^8+3r^10}{2d^8r^4(4\pi+\zeta)}.
\end{equation}
We combine Eqs.\eqref{g53i} and \eqref{g53c} to obtain
\begin{equation}\label{g53d}
\frac{\{r^3(d^2-r^2)^3(d^2+3r^2)\}
A_2'}{\{r^4(d^2-r^2)^3(r^2-d^2)\}A_2} +\frac{A_2''}{A_2}=0.
\end{equation}
Solving this second-order differential equation yields the
expression for $A_2$ as
\begin{equation}\label{g53h}
A_2(r)=\frac{C_1}{2(d^2-r^2)}+C_2.
\end{equation}
The integration constants $C_1$ and $C_2$ shall be determined using
boundary conditions. The metric potentials from Eqs.\eqref{g52} and
\eqref{g53h} can be used to rewrite the field equations
\eqref{g18}-\eqref{g20} as follows
\begin{align}\nonumber
\rho&=\frac{1}{12 d^8 r^4 (\zeta +2 \pi ) (\zeta +4 \pi )\{2 R
(d^2-r^2)+C_1\}}[C_1\{-d^8 (5 \zeta +12 \pi )\\\nonumber& \times
q^2+18 d^6 (3 \zeta +8 \pi )r^4-6 d^4 (23 \zeta +60 \pi ) r^6 +2 d^2
(65 \zeta +168 \pi ) r^8 \\\nonumber &-6(7 \zeta +18 \pi )
r^{10}\}-2 C_2(d^2-r^2)\{d^8(5\zeta +12 \pi ) q^2-48 d^6 (\zeta +3
\pi )\\\label{15} & \times r^4+120 d^4 (\zeta +3 \pi ) r^6-112 d^2
(\zeta +3 \pi ) r^8+36 (\zeta +3\pi)r^{10}\}],
\\\nonumber p_r&=\frac{1}{12 d^8 r^4 (\zeta +2 \pi )
(\zeta +4\pi )\{2 C_2(d^2-r^2)+C_1\}}[C_1\{d^8 (5 \zeta +12 \pi)
q^2\\\nonumber& +18 d^6 \zeta  r^4-6 d^4 (13 \zeta +12 \pi) r^6+2
d^2 (43 \zeta +48 \pi) r^8-6 (5 \zeta +6 \pi )r^{10}\}\\\nonumber
&+2 C_2(d^2-r^2)\{d^8 (5 \zeta +12 \pi )q^2-48 \pi d^6 r^4+24 d^4
(3\pi-\zeta) r^6+16 d^2\\\label{16} & \times (2 \zeta -3 \pi )r^8+12
(\pi -\zeta ) r^{10}\}],
\\\nonumber p_t&=\frac{1}{12 d^8 r^4 (\zeta +2 \pi )
(\zeta +4 \pi )\{2C_2(d^2-r^2)+C_1\}}[C_1\{-d^8(7\zeta +12 \pi
)\\\nonumber&\times q^2+18 d^6 \zeta r^4-42 d^4 \zeta r^6+38
d^2\zeta r^8-12 \zeta r^{10}\}-2 C_2
(d^2-r^2)\{d^8\\\nonumber&\times (7 \zeta +12 \pi)q^2+48 \pi
d^6r^4-12 d^4 (\zeta +12 \pi) r^6+16 d^2 (\zeta +9
\pi)r^8-6\\\label{17}&\times (\zeta +8\pi )r^{10}\}].
\end{align}
The above system still includes an unknown function, i.e., an
electric charge. Its known exponential form has been used as
$q=\alpha r^3$, where $\alpha$ is a constant with a dimension of
$\ell^{-2}$ \cite{ag}.

\section{Closed-form Solution of Additional Fluid Content}

This section considers an additional source of matter distribution,
which adds uncertainty to the field equations. To solve these
equations, we apply two distinct limitations, as mentioned in the
subsections presented below.

\subsection{Model I: Density-like Constraint}

To calculate the total internal fluid solution, we use a condition
based on the energy density of both main and secondary sources. The
constraint is specified as \cite{82}
\begin{align}\label{z2}
\rho^*=\Psi_{0}^{0},
\end{align}
where $\rho^*$ indicates the effective energy density whose
mathematical expression is given by the left side of
Eq.\eqref{g8aa}. We substitute this with Eq.\eqref{g21} into
\eqref{z2}, yielding the differential equation as
\begin{align}\label{z3}
-\frac{1}{r^2 B_2^2}+\frac{2 B_2'}{r
B_2^3}+\frac{1}{r^2}+\frac{q^2}{r^4}=\frac{2 \bar{t}}{r
B_2^3}\bigg(\frac{3 B_2'}{B_2}-\frac{1}{r}\bigg)-\frac{2 \bar{t}'}{r
B_2^3}.
\end{align}
The deformation function $\bar{t}(r)$ is the only unknown in the
above equation. This function is computed as
\begin{align}\nonumber
\bar t&=\frac{1}{60r(r^2-d^2)^5}\big[15(d-r)(d+r)\{(3\alpha^2
d^{13}+d^9)\tanh^{-1}\big(r/d\big)-4 C_3\}\\\label{z4}&+d^4r\{-15
d^6+130 d^4 r^2-154 d^2 r^4+15\alpha^2 d^8(2 r^2-3
d^2)+54r^6\}\big],
\end{align}
where $C_3$ is the integration constant. Equation \eqref{z4} is thus
utilized to calculate the components of the newly defined source
$\Psi_{\alpha\eta}$, as follows
\begin{align}\label{z5}
\Psi_{0}^{0}&=-\frac{\alpha ^2 d^8 r^2-12 d^6+30 d^4 r^2-28 d^2
r^4+9 r^6}{8 \pi d^8},
\\\nonumber \Psi_{1}^{1}&=\frac{1}{240\pi d^{12}r^3\{2 C_2 (d^2-r^2)+C_1\}}
\big[\{2C_2(d^2-r^2)^2+C_1(d^2-5
r^2)\}\\\nonumber&\times\{d^4\{-45\alpha^2 d^{10}r+30\alpha^2 d^8
r^3-15 d^6 r+130 d^4 r^3-154 d^2 r^5+15\\\label{z6}& \times
d^5(3\alpha^2 d^4+1)(d^2-r^2)\tanh^{-1}\big(r/d\big)+54 r^7\}-60
C_3(d^2-r^2)\}\big],
\\\nonumber \Psi_{2}^{2}&=-\frac{1}{480 \pi d^{12} r^3(d^2-r^2)\{2 C_2 (d^2-r^2)+C_1\}}
\big[2 C_2(d^2-r^2)^2\{d^4\{15d^5\\\nonumber&\times(3\alpha^2
d^4+1)(d^4-22 d^2 r^2+21 r^4) \tanh^{-1}\big(r/d\big)-r(45\alpha^2
d^{12}-975\\\nonumber&\times \alpha^2d^{10} r^2+15 d^8(40\alpha^2
r^4+1)-85 d^6 r^2+1984 d^4 r^4-2448 d^2 r^6+9 \\\nonumber& \times 96
r^8)\}-60 C_3(d^4-22 d^2 r^2+21 r^4)\}+C_1(d^2+r^2)\{d^4\{15 d^5
(3\alpha^2 d^4\\\nonumber & +1)(d^4-14 d^2 r^2+13
r^4)\tanh^{-1}\big(r/d\big)-r \{45 \alpha^2 d^{12}-615\alpha^2d^{10}
r^2+5 \\\nonumber& \times 3d^8(24\alpha^2 r^4+1)+35 d^6 r^2+944 d^4
r^4-1216 d^2 r^6+432 r^8\}\}-60 C_3\\\label{z7}& \times(d^4-14 d^2
r^2+13 r^4)\}\big].
\end{align}
Using Eqs.\eqref{15}-\eqref{17} and \eqref{z5}-\eqref{z7}, the
expressions in Eq.\eqref{g13} provide the complete solution for
mimicking density constraint.

Junction conditions are important for compact astrophysical objects
as they connect physical fields and their rates of change across
boundaries. To ensure a smooth transition between connected
geometries of spacetime, both the metric tensor and extrinsic
curvature must be continuous. This also assures that the generic
relativistic field equations are consistent. These conditions are
critical for resolving boundary problems, analyzing wave
propagation, and studying material interactions. We use the
Reissner-Nordstr\"{o}m spacetime to depict geometry outside the
object, as given below
\begin{equation}\label{18}
ds^2=-\bigg(1-\frac{2M}{r}+\frac{Q^2}{r^2}\bigg)dt^2
+\bigg(1-\frac{2M}{r}+\frac{Q^2}{r^2}\bigg)^{-1}dr^2+r^2d\Omega^2,
\end{equation}
where $r > 2M$, with $M$ and $Q$ being the total mass and charge of
the star. The conditions are divided into two forms \cite{41ja}. To
achieve a smooth shape, the first step involves maintaining
consistent values of the metric tensor along the boundary. The
second component connects spacetime sectors by preserving a
continuous extrinsic curvature. At the stellar boundary, which is
defined as $r=R$, radial pressure equals to zero. These conditions
result in the following three mathematical expressions as
\begin{align}\label{18a}
1-\frac{2M}{R}+\frac{Q^2}{R^2}&=\bigg(\frac{d^2-R^2}{d^2}\bigg)^4,\\\label{18b}
\sqrt{1-\frac{2M}{R}+\frac{Q^2}{R^2}}&=\frac{C_1}{2(d^2-R^2)}+C_2,\\\nonumber
\bar p_r(R)=0&=\frac{1}{(\zeta +2\pi)(\zeta +4 \pi)}[20
d^4\{C_1\{d^8 (5 \zeta +12 \pi ) Q^2+18 d^6 \\\nonumber&\times\zeta
R^4-6 d^4 (13 \zeta +12 \pi ) R^6+2 d^2 (43 \zeta +48 \pi ) R^8-6
\\\nonumber&\times(5\zeta +6 \pi ) R^{10}\}+2 C_2(d^2-R^2)\{d^8 (5\zeta+12
\pi) Q^2-48\\\nonumber&\times \pi d^6 R^4+24 d^4 (3 \pi -\zeta)
R^6+16 d^2(2 \zeta-3 \pi ) R^8+12 (\pi\\\nonumber&-\zeta)
R^{10}\}\}]+\frac{1}{\pi}[R \omega\{2 C_2(d^2-R^2)^2+C_1(d^2-5
R^2)\}\{d^4\\\nonumber&\times(-45 \alpha^2 d^{10} R+30 \alpha ^2 d^8
R^3-15 d^6 R+130 d^4R^3-154 d^2\\\nonumber&\times R^5+15 d^5(3
\alpha^2d^4+1)(d^2-R^2)\tanh^{-1}\big(r/d\big)+54R^7)\\\label{18c}&-60
C_3 (d^2-R^2)\}].
\end{align}
Simultaneously solving the above equations yields the following
expressions for ($d^2,C_1,C_2$) as
\begin{align}\label{19}
d^2&=\frac{R^2}{1-\bigg(\frac{Q^2-2 M R+R^2}{R^2}\bigg)^{1/4}},
\\\nonumber C_1&=\frac{(\zeta+2
\pi)(\zeta+4\pi)(d^2-R^2)}{2
R^3\{d^4(\zeta^2+6\pi\zeta+8\pi^2)\phi-60
C_3(\zeta^2+6\pi\zeta+8\pi^2)\omega(d^2-R^2)\}}\\\nonumber
&\times\sqrt{\frac{-2 M
R+Q^2+R^2}{R^2}}\bigg[R\omega(d^2-R^2)\big\{d^4(-45\alpha^2
d^{10}R+6 5\alpha^2 d^8 R^3\\\nonumber &-15 d^6 R+130 d^4 R^3-154
d^2R^5+54 R^7+15d^5(3 \alpha^2 d^4+1)(d^2-R^2)\\\nonumber
&\times\tanh^{-1}\big(R/d\big))-60 C_3
(d^2-R^2)\big\}+\frac{\pi}{(\zeta+2\pi)(\zeta +4\pi)}\big\{20
d^4\{d^8(5\zeta\\\nonumber & +12\pi)Q^2-48\pi d^6 R^4+24 d^4 (3 \pi
-\zeta )R^6+16 d^2(2\zeta-3\pi) R^8\\\label{20} &+12(\pi-\zeta)
R^{10}\}\big\}\bigg],
\\\nonumber C_2&=\sqrt{\frac{-2 M R+Q^2+R^2}{4R^2}}\bigg[\frac{(\zeta
+2\pi)(\zeta+4\pi)}{2 R^3\{d^4\phi-60 C_3(\zeta^2+6 \pi
\zeta+8\pi^2)\omega(d^2-R^2)\}}\\\nonumber & \times\{R
\omega(d^2-R^2)\{d^4(-45\alpha^2 d^{10} R+30\alpha^2 d^8 R^3-15
d^6R+130 d^4 R^3-154 \\\nonumber &\times d^2 R^5+15 d^5(3\alpha^2
d^4+1)(d^2-R^2)\tanh^{-1}\big(R/d\big)+54R^7)-60 C_3
(d^2\\\nonumber& -R^2)\}\}+\frac{1}{(\zeta+2\pi)(\zeta +4\pi)}\{20
d^4\{d^8(5\zeta+12\pi)Q^2-48\pi d^6 R^4+24 d^4\\\label{21}& \times
(3\pi-\zeta)R^6+16 d^2(2 \zeta-3\pi)R^8+12(\pi-\zeta)
R^{10}\}\}\bigg],
\end{align}
where
\begin{align}\nonumber
\phi&=15 d^5 (\zeta^2+6\pi\zeta+8\pi^2)\omega(3\alpha^2
d^4+1)(d^2-R^2)\tanh^{-1}\big(R/d\big)+R\\\nonumber&\times\{-45\alpha^2
d^{10}(\zeta^2+6\pi\zeta+8\pi^2)\omega+30\alpha^2 d^8\theta
R^2\omega-15
d^6\{\zeta^2\omega+6\pi\zeta\\\nonumber&\times(\omega+1)+8
\pi^2(\omega+2)\}+10 d^4 R^2\{13\zeta^2 \omega+3\pi\zeta(26 \omega
+9)+8 \pi^2 \\\nonumber&\times(13 \omega+9)\}-2 d^2 R^4\{77\zeta^2
\omega+3\pi\zeta(154 \omega+45)+8\pi^2(77\omega+45)\}\\\nonumber&+6
R^6\{9\zeta^2\omega+3\pi\zeta(18 \omega+5)+8\pi^2(9 \omega+5)\}\}.
\end{align}
These values will be used to produce a graphical analysis of our
solution. By incorporating the parameter $\omega$ and $\zeta$ into
these constants allows a better understanding of how changes in the
action affect the stellar properties.

\subsection{Model II: Pressure-like Constraint}

We now construct a second model using the radial pressure
restriction as \cite{83}
\begin{align}\label{z11}
p_r^*=\Psi_{1}^{1},
\end{align}
where $p_r^*$ indicates the effective radial pressure whose
mathematical expression is given by the left side of
Eq.\eqref{g9aa}. The following equation is obtained by substituting
this choice along with Eq.\eqref{g22} into \eqref{z11} as
\begin{align}\label{z12}
\frac{2 A_2'}{r A_2 B_2^2}+\frac{1}{r^2
B_2^2}-\frac{1}{r^2}+\frac{q^2}{r^4}=\frac{2 \bar t}{r
B_2^3}\bigg(-\frac{2 A_2'}{A_2}-\frac{1}{r}\bigg).
\end{align}
The function $\bar t(r)$ can be directly obtained from the above
algebraic equation as
\begin{align}\nonumber
\bar t&=\frac{1}{4 C_2(d^2-r^2)^7+2 C_1(d^2-5 r^2)(d^2-r^2)^5}[d^4
r^2\{C_1(6 d^4 r^2\\\label{z13}&-8 d^2 r^4+3 r^6)+2 C_2(4 d^8-10 d^6
r^2+10 d^4 r^4-5 d^2 r^6+r^8)\}],
\end{align}
which helps to compute the components of the newly defined source as
follows
\begin{align}\nonumber
\Psi_{0}^{0}&=\frac{1}{8 \pi d^8(2 C_2(d^2-r^2)^2+C_1(d^2-5
r^2))^2}[-4 C_2^2(12 d^8-26 d^6 r^2\\\nonumber&+34 d^4 r^4-21 d^2
r^6+5 r^8)(d^2-r^2)^3-4 C_1 C_2(6 d^8-10 d^6 r^2\\\nonumber&+34 d^4
r^4-37d^2 r^6+13 r^8)(d^2-r^2)^2+C_1^2(-30 d^8 r^2+164 d^6
r^4\\\label{z14}&-297 d^4 r^6+246 d^2 r^8-75 r^{10})],
\\\nonumber \Psi_{1}^{1}&=\frac{1}{8 \pi d^8\{2C_2(d^2-r^2)+C_1\}}[C_1(-6 d^4 r^2+8 d^2 r^4-3 r^6)
\\\label{z15}&-2 C_2(4 d^8-10 d^6 r^2+10 d^4 r^4-5 d^2 r^6+r^8)],
\\\nonumber \Psi_{2}^{2}&=\frac{1}{2 \pi  d^8
(d^2-r^2)\{2 C_2(d^2-r^2)+C_1\}\{2 C_2(d^2-r^2)^2+C_1(d^2-5
r^2)^2\}}\\\nonumber&\times[-8 C_2^3(d^2-r^2)^9+C_1^3 d^2 r^2(-3
d^8+9 d^6 r^2+4 d^4 r^4-6 d^2 r^6+2 r^8)-8\\\nonumber&\times C_1
C_2^2(d^{10}-d^8 r^2+8 d^6 r^4-12 d^4 r^6+8 d^2 r^8-2
r^{10})(d^2-r^2)^3-2 C_1^2\\\nonumber&\times C_2(d^{14}+4 d^{12}
r^2-29 d^{10} r^4+80 d^8 r^6-145 d^6 r^8+153 d^4 r^{10}-81 d^2
r^{12}\\\label{z16}&+17 r^{14})].
\end{align}
Using Eqs.\eqref{15}-\eqref{17} and \eqref{z14}-\eqref{z16}, the
expressions in \eqref{g13} gives a complete solution for mimicking
pressure constraint.

Darmois junction conditions are again needed at this stage to find
the junction conditions. Continuing the first and second fundamental
forms yield the following equations as
\begin{align}\label{18aa}
1-\frac{2M}{R}+\frac{Q^2}{R^2}&=\bigg(\frac{d^2-R^2}{d^2}\bigg)^4,\\\label{18bb}
\sqrt{1-\frac{2M}{R}+\frac{Q^2}{R^2}}&=\frac{C_1}{2(d^2-R^2)}+C_2,\\\nonumber
\bar p_r(R)&=\frac{1}{(\zeta
+2\pi)(\zeta+4\pi)R^4}[2\{C_1(d^8(5\zeta+12\pi)Q^2+18 d^6\zeta
\\\nonumber & \times R^4-6 d^4(13\zeta+12\pi)R^6+2
d^2(43\zeta+48\pi)R^8-6(5\\\nonumber &\times\zeta+6\pi) R^{10})+2
C_2(d^2-R^2)\{d^8(5\zeta+12\pi)Q^2-48\\\nonumber &\times \pi d^6
R^4+24 d^4 (3 \pi-\zeta) R^6+16 d^2(2\zeta-3\pi)R^8+12\\\nonumber
&\times(\pi-\zeta) R^{10}\}\}]+\frac{1}{\pi}[3 \omega\{C_1
R^2(\alpha^2 d^8-6 d^4+8 d^2 R^2\\\nonumber&-3 R^4)+2
C_2(d^2-R^2)(\alpha^2 d^8R^2-4 d^6+6 d^4 R^2-4 d^2
\\\label{18cc}&\times R^4+R^6)\}]=0.
\end{align}
The above system of equations is solved to get the following
expressions for the constants ($C_1, C_2$) as
\begin{align}\nonumber
C_1&=-\frac{1}{6 R^4 (d^2-R^2)^2 \{(\zeta+2\pi)(\zeta+4
\pi)\omega+\pi(3 \zeta+8 \pi)\}}\\\nonumber &\times\sqrt{\frac{-2 M
R+Q^2+R^2}{R^2}}\big[2 \pi(d^8 (5 \zeta+12 \pi) Q^2-48 \pi
\\\nonumber &\times d^6 R^4+24 d^4 (3 \pi -\zeta) R^6+16 d^2 (2
\zeta-3 \pi) R^8+12 \\\nonumber& \times(\pi-\zeta) R^{10})+3
(\zeta+2 \pi)(\zeta+4 \pi) R^4 \omega(\alpha^2 d^8
R^2-4\\\label{19a}& \times d^6+6 d^4 R^2-4d^2
R^4+R^6)\big],\\\nonumber C_2&=\frac{1}{12
R^4(R^2-d^2)^3\{(\zeta+2\pi)(\zeta+4\pi)
\omega+\pi(3\zeta+8\pi)\}}\\\nonumber&\times\sqrt{\frac{-2 M
R+Q^2+R^2}{R^2}}\big[3(\zeta+2\pi)(\zeta+4\pi)R^6\omega(-\alpha^2
\\\nonumber&\times d^8+6 d^4-8 d^2 R^2+3 R^4)+2\pi(-d^8
(5\zeta+12\pi)Q^2\\\nonumber&-18 d^6\zeta R^4+6
d^4(13\zeta+12\pi)R^6-2 d^2 (43\zeta+48\pi) R^8\\\label{20a}&+6
(5\zeta+6\pi)R^{10})\big],
\end{align}
where $d^2$ has the same value as obtained in Eq.\eqref{19}. These
values will be used to graphically examine the resulting model.

\section{Analyzing Physical Feasibility Criteria}

This section contains an extensive evaluation of the proposed
models. Their internal structural properties will be the primary
focus of this research. This study examines the stability of the
solutions under different pressures and environmental conditions.
The usefulness and reliability of solutions depend on their metric
functions and matter properties. We use two and five distinct
decoupling and model parameter values as $\omega = 0.1, 0.2$ and
$\zeta=0.2,0.4,0.6,0.8$ together with $Q=0.3, 0.6$ to the star LMC
X-4 \cite{89fa}. This candidate has a mass $M=1.04 \pm
0.09~M_{\bigodot}$ and radius $R=8.301 \pm 0.2~km$. A graphical
analysis of the stability and structure of compact stars is based on
this data.

\subsection{Metric Elements and Profile of Fluid Triplet}

To be physically viable, a star model should not have geometric
singularities. To ensure metric potentials in the stellar interior,
it is important to study their behavior. To create a physically
realistic interior, these components should be finite at the core
and gradually increase towards the surface. The $g_{rr}$ and
$g_{tt}$ potentials should have a value of $1$ and remain between
$0$ and $1$ in the center, respectively. Additionally, the
derivatives of these values must be zero at the origin. Figure
\textbf{1} shows how the four metric components interact positively
and smoothly within the star.
\begin{figure}[h!]\center
\epsfig{file=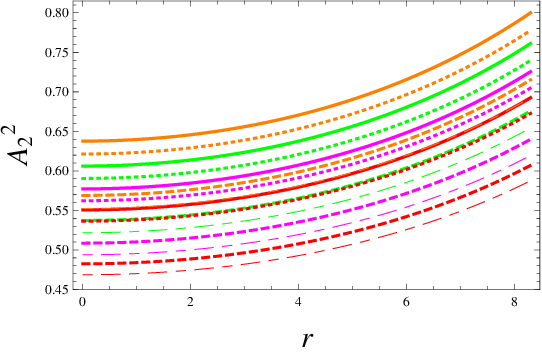,width=0.45\linewidth}\epsfig{file=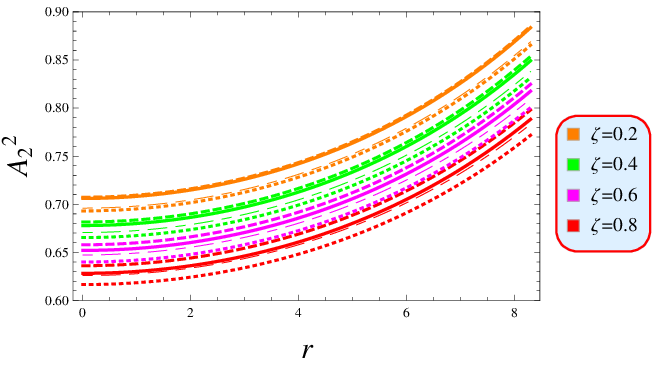,width=0.54\linewidth}
\epsfig{file=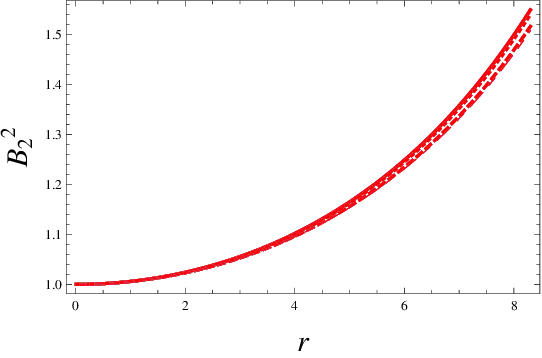,width=0.45\linewidth}\epsfig{file=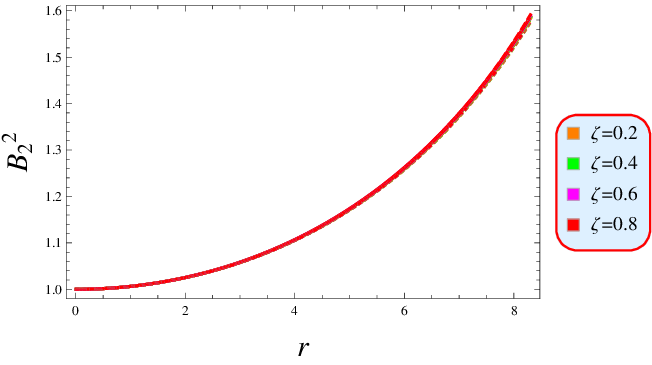,width=0.54\linewidth}
\caption{Metric elements for solutions 1 (left) and 2 (right) for
$Q=0.3$ at $\omega=0.1$ (thick) \& $0.2$ (dashed), and for $Q=0.6$
at $\omega=0.1$ (dotted) \& $0.2$ (small-long dashed).}
\end{figure}

Understanding fluid variables is a key to understanding star
formation and equilibrium. The star's energy density must be
positive and limited, gradually decreasing from center to surface.
Radial pressure, which radiates at right angles, is higher at the
center and progressively declines to zero on the outer surface.
Anisotropy occurs when tangential and radial pressures have distinct
effects on a surface. Figure \textbf{2} demonstrates consistent
patterns in both solutions. The second model produces a denser star,
as compared to the first. Both models show similar radial and
tangential pressure patterns. It is also shown that at the boundary
$r=R$, the radial pressure is vanished. Anisotropy, whether positive
or negative, significantly affects a stellar model's stability.
Positive anisotropy means a star's radial pressure is lower than its
tangential pressure, leading to increased stability. This outward
push counteracts gravity and prevents the star from collapsing.
Negative anisotropy causes radial pressure to exceed tangential
pressure, resulting in decreased outward support and instability.
Negative anisotropy causes stars to remain stable for shorter
periods compared to positive anisotropy. Figure \textbf{2}
demonstrates that anisotropy has a bigger impact on the outside
layers than the center, resulting in a repulsive force that promotes
systems' stability.
\begin{figure}[h!]\center
\epsfig{file=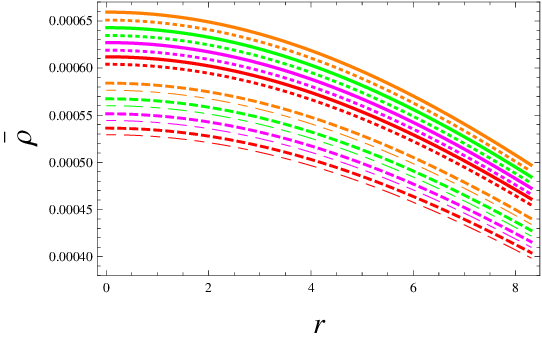,width=0.45\linewidth}\epsfig{file=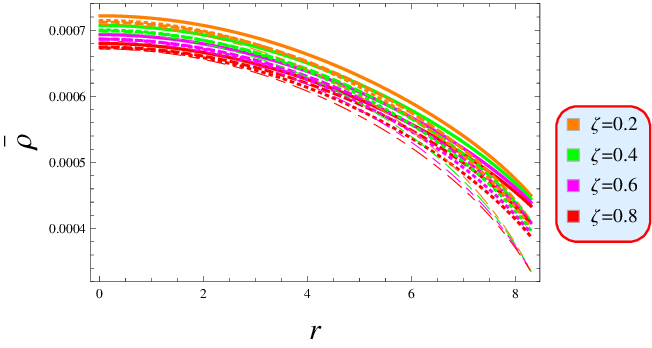,width=0.54\linewidth}
\epsfig{file=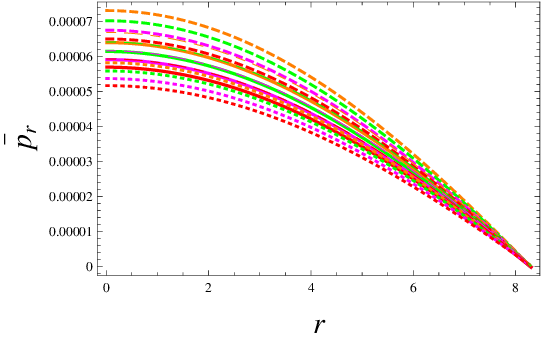,width=0.45\linewidth}\epsfig{file=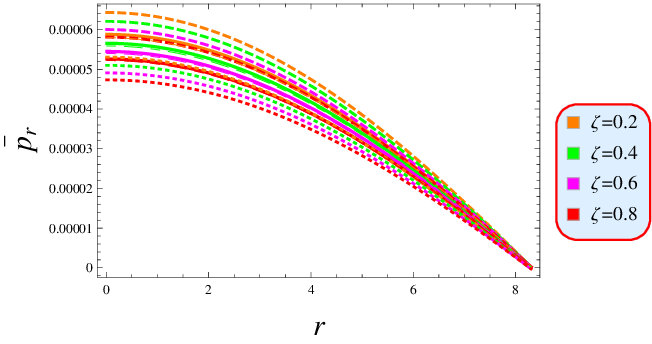,width=0.54\linewidth}
\epsfig{file=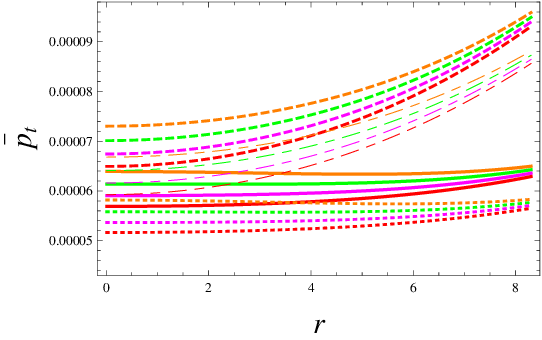,width=0.45\linewidth}\epsfig{file=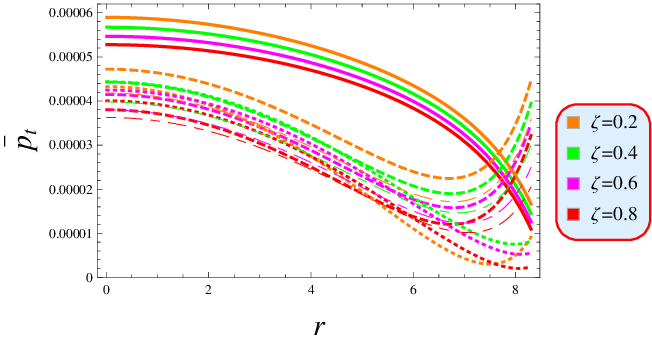,width=0.54\linewidth}
\epsfig{file=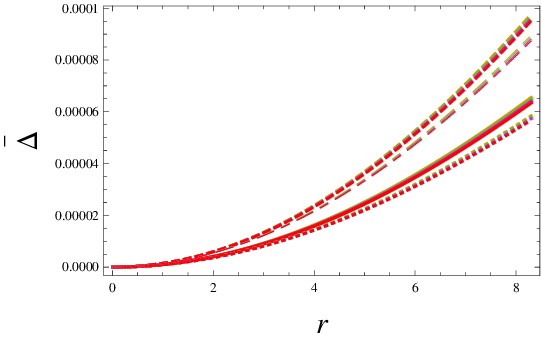,width=0.45\linewidth}\epsfig{file=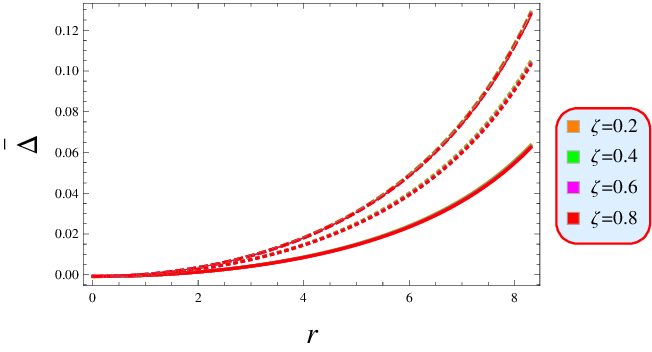,width=0.55\linewidth}
\caption{Fluid parameters for models 1 (left) and 2 (right) for
$Q=0.3$ at $\omega=0.1$ (thick) \& $0.2$ (dashed), and for $Q=0.6$
at $\omega=0.1$ (dotted) \& $0.2$ (small-long dashed).}
\end{figure}

The validity of the solution is based on the rate of change in
physical quantities with the radius. This relationship may be
expressed as
$$\bar\rho'|_{r=0}=0=\bar p_r'|_{r=0}, \quad \bar\rho''|_{r=0}<0, \quad \bar
p_r''|_{r=0}<0.$$ We have verified the consistent behavior of both
solutions.

\subsection{Mass and Other Physical Components}

The mass function can be determined both analytically or
numerically, depending on the energy density. Figure \textbf{3} (top
graphs) illustrates that the mass function decreases as $r$
approaches zero. Beyond this point, the function gradually grows as
the radial distance increases.

The mass-to-radius ratio of an object determines its compactness,
which impacts its gravitational energy and stability. Neutron stars
have strong gravitational fields due to their compactness. According
to Buchdahl \cite{42a}, a stable compact object must satisfy the
requirement $\big(\mu(r)=\frac{m(r)}{r} < 0.44\big)$, which
indicates dimensionless compactness. The limit was initially defined
for a spherical fluid with uniform pressure. According to research,
it holds true even when internal pressures change. Light is forced
to leave a big/massive object by gravity, which results in energy
loss. Light waves stretch and increase in wavelength as moving far
away from gravitational sources. This is known as surface redshift,
and it is expressed as
\begin{equation}
z=\bigg\{1-2\bigg(\frac{m(r)}{r}\bigg)\bigg\}^{-\frac{1}{2}}-1.
\end{equation}
Internal redshift increases with increasing compactness. A maximum
surface redshift of $5.211$ can be produced by anisotropic matter
configurations. Achieving this limit is crucial for correct modeling
in anisotropic systems \cite{89g}. Figure \textbf{3} shows
consistent trends and acceptable behavior throughout the domain.
\begin{figure}[h!]\center
\epsfig{file=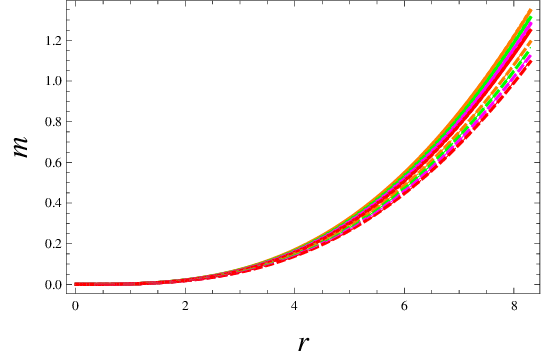,width=0.45\linewidth}\epsfig{file=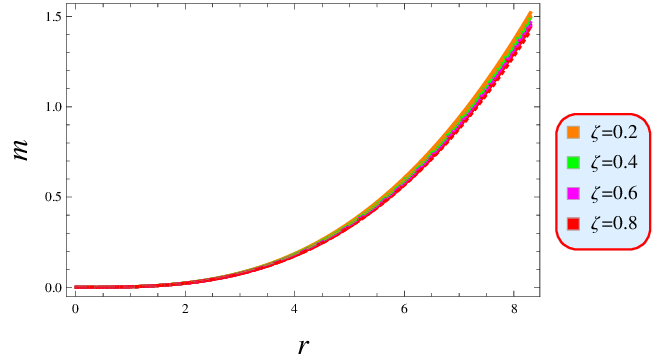,width=0.54\linewidth}
\epsfig{file=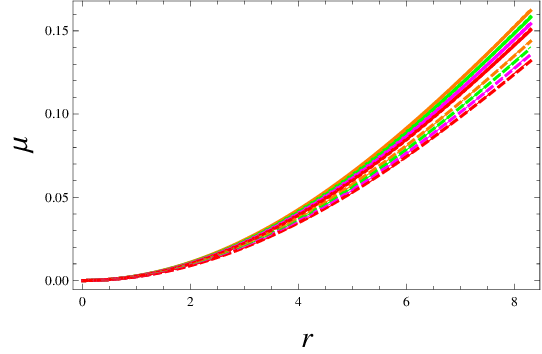,width=0.45\linewidth}\epsfig{file=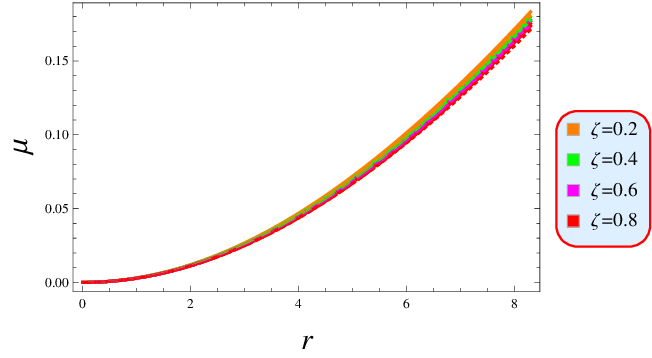,width=0.54\linewidth}
\epsfig{file=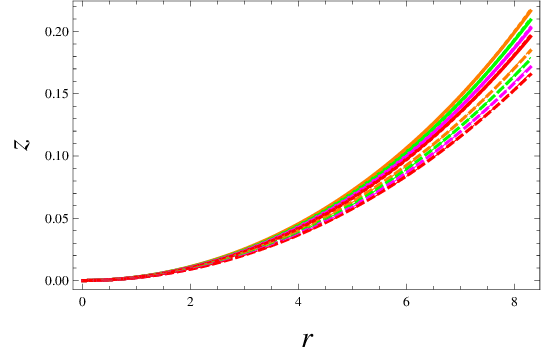,width=0.45\linewidth}\epsfig{file=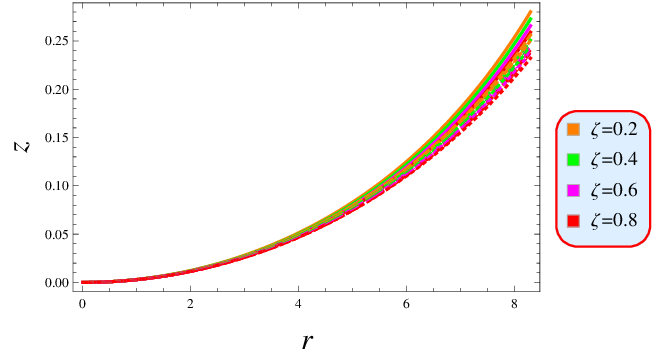,width=0.54\linewidth}
\caption{Physical quantities for models 1 (left) and 2 (right) for
$Q=0.3$ at $\omega=0.1$ (thick) \& $0.2$ (dashed), and for $Q=0.6$
at $\omega=0.1$ (dotted) \& $0.2$ (small-long dashed).}
\end{figure}

\subsection{Rotational Dynamics}

Researchers measure the moment of inertia of celestial objects to
understand their rotation. Astronomers use this metric to study the
distribution of mass in dense star configurations and its impact on
spin. Additionally, it connects the body's stationary features with
its rotating dynamics. An object with a constant mass and fixed
radius has the highest moment of inertia when its rotational speed
reaches its lowest possible value as given in \cite{90}
\begin{equation}
I=0.4\left[\left(1+\frac{M}{R}\right)M R^2\right].
\end{equation}
Figure \textbf{4} illustrates how higher mass increases the moment
of inertia. Adding mass to an object improves its resistance to
angular acceleration by moving it away from the rotating axis.
Rotational behavior is constant when the object's resistance to
rotation reaches the maximum level for its mass. When a body exceeds
this limit, adding more mass does not significantly affect its
rotation.
\begin{figure}[h!]\center
\epsfig{file=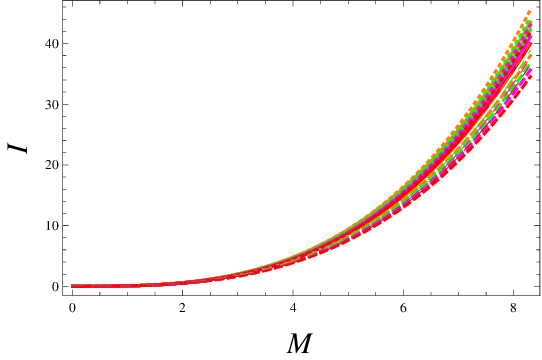,width=0.45\linewidth}\epsfig{file=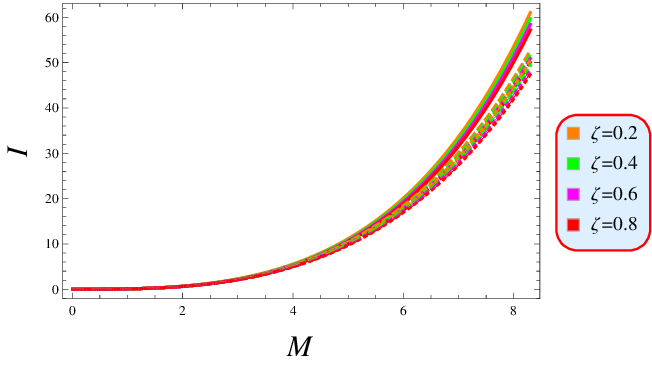,width=0.54\linewidth}
\caption {Rotational dynamics for solutions 1 (left) and 2 (right)
for $Q=0.3$ at $\omega=0.1$ (thick) \& $0.2$ (dashed), and for
$Q=0.6$ at $\omega=0.1$ (dotted) \& $0.2$ (small-long dashed).}
\end{figure}

\subsection{Equation of State}

The equation of state is given by $\bar{p}_i = \omega_i \bar{\rho}$,
where $\omega_i$ determines radial and tangential parameters for
$i=r,t$. A physical fluid distribution requires a ratio of pressure
components to energy density between $0$ and $1$. Meeting this
criteria assures solutions adhere to the basic thermodynamic rules
of star formation. Figure \textbf{5} demonstrates that the
transverse and radial pressure to density ratios are less than $1$,
suggesting that energy density is much larger than pressures in the
gravitational systems.
\begin{figure}[h!]\center
\epsfig{file=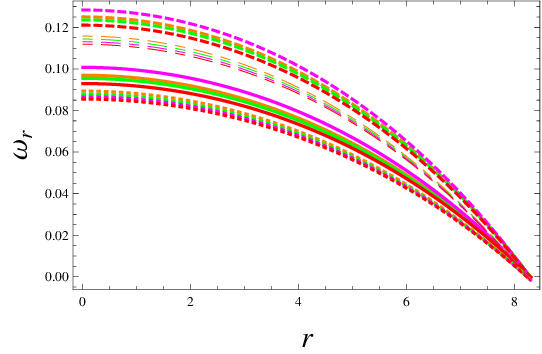,width=0.45\linewidth}\epsfig{file=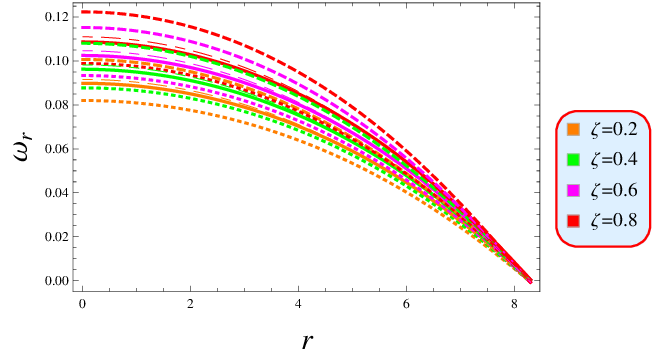,width=0.54\linewidth}
\epsfig{file=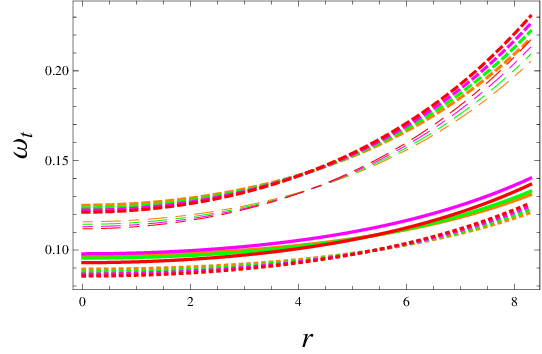,width=0.45\linewidth}\epsfig{file=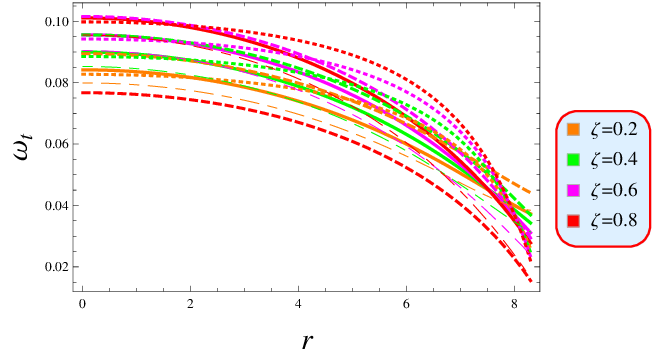,width=0.54\linewidth}
\caption{Realistic parameters for solutions 1 (left) and 2 (right)
for $Q=0.3$ at $\omega=0.1$ (thick) \& $0.2$ (dashed), and for
$Q=0.6$ at $\omega=0.1$ (dotted) \& $0.2$ (small-long dashed).}
\end{figure}

\subsection{Viability Analysis}

The internal configuration of a star model must satisfy energy
requirements in order to be physically feasible. The $null$, $weak$,
$dominant$, $strong$, and $trace$ energy conditions are the five
categories into which these conditions fall. They are essential for
distinguishing between regular and exotic matter when analyzing
matter composition in a theoretical framework. Enforcing these
limits allows researchers to set appropriate bounds for fluid
variables, to make sure that model results correspond to actual
data. The energy conditions are expressed as

$Null: \bar\rho+\bar p_t+\frac{q^2}{4\pi r^4}\geq0, \quad
\bar\rho+\bar p_r\geq0$,

$Weak: \bar\rho+\bar p_t+\frac{q^2}{4\pi r^4}\geq0, \quad
\bar\rho+\bar p_r\geq0, \quad \bar\rho+\frac{q^2}{8\pi r^4}\geq0$,

$Dominant: \bar\rho+\bar p_t+\frac{q^2}{4\pi r^4}\geq0,
\quad\bar\rho-\bar p_t\geq0, \quad \bar\rho-\bar p_r+\frac{q^2}{4\pi
r^4}\geq0, \quad\bar\rho+\bar p_r\geq0$,

$Strong: \bar\rho+\bar p_r+2\bar p_t+\frac{q^2}{4\pi r^4}\geq0,
\quad \bar\rho+\bar p_t+\frac{q^2}{4\pi r^4}\geq0, \quad
\bar\rho+\bar p_r\geq0$,

$Trace: \bar\rho-\bar p_r-2\bar p_t\geq0$.

\noindent The viability of solution can be determined by analyzing
only the $dominant$ and $trace$ energy conditions as all physical
parameters have positive values, as shown in Figure \textbf{2}.
Figure \textbf{6} shows that both models consistently exhibit
positive behavior under the given conditions. Positive results
across the domain indicate the presence of ordinary matter in star
interiors, supporting the physical acceptability of these solutions.
\begin{figure}[h!]\center
\epsfig{file=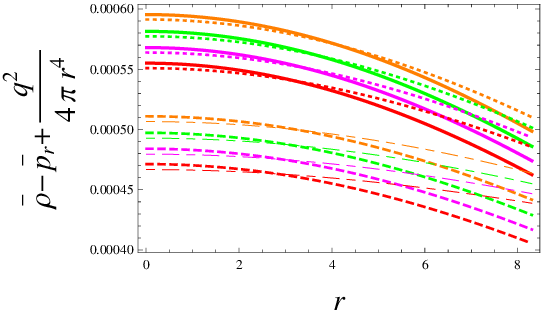,width=0.47\linewidth}\epsfig{file=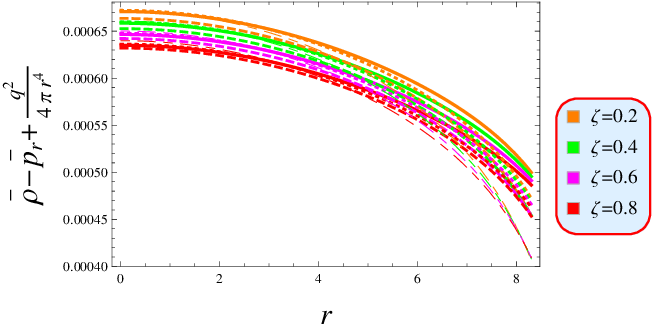,width=0.54\linewidth}
\epsfig{file=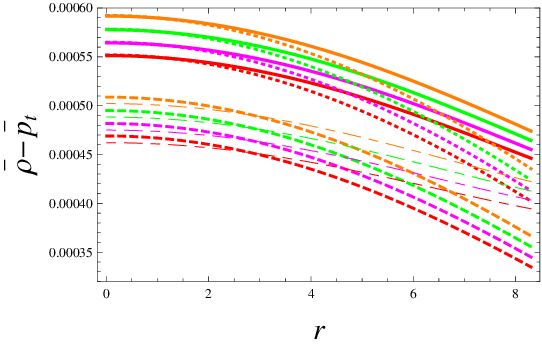,width=0.45\linewidth}\epsfig{file=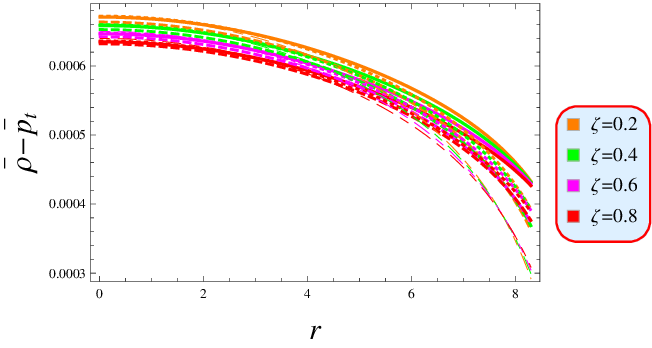,width=0.54\linewidth}
\epsfig{file=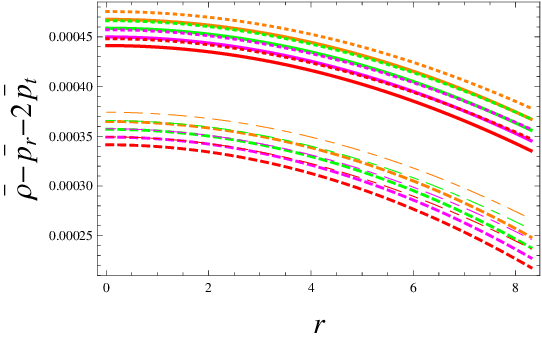,width=0.45\linewidth}\epsfig{file=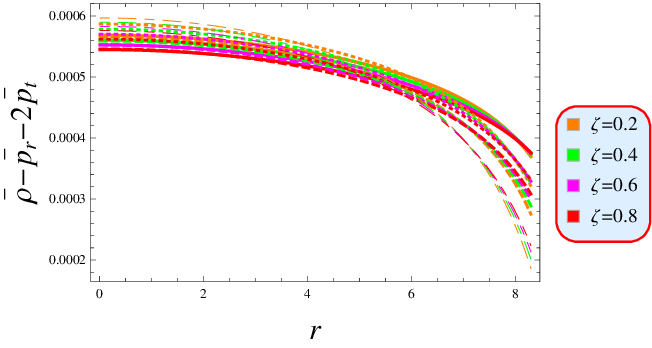,width=0.54\linewidth}
\caption{Viability criteria for models 1 (left) and 2 (right) for
$Q=0.3$ at $\omega=0.1$ (thick) \& $0.2$ (dashed), and for $Q=0.6$
at $\omega=0.1$ (dotted) \& $0.2$ (small-long dashed).}
\end{figure}

\subsection{Stability Analysis}

Understanding the nature of celestial objects requires research into
their structural stability. Researchers examine how compact stars
react to disruptive forces and maintain stability. Researchers
utilize a tool to evaluate star stability and anticipate collapse.
There are numerous analytical and computational techniques
available. The following subsections provide an overview of the key
techniques used in critical analysis.

\subsubsection{Equilibrium of Forces}

Several factors affect the system's internal structure, disturbing
its hydrostatic equilibrium. To confirm the equilibrium of a compact
model, five forces are evaluated, gravitational, anisotropic,
electromagnetic, modified and hydrostatic. The validation principle
states that the net force is equal to zero, which is expressed as
\begin{eqnarray}\label{hg52}
f_g+f_h+f_a+f_m+f_{e}=0.
\end{eqnarray}
For dense stellar bodies with anisotropy, the evolution equation
describes their interior structure as follows \cite{42c,42d}
\begin{eqnarray}\label{hhg52}
-\frac{M_G}{r}(\bar\rho+\bar p_r)\frac{A_2}{B_2}-\frac{2}{r}(\bar
p_r-\bar p_t)-\bar p_r'-\frac{\alpha(\bar p_r'+2\bar
p_t'+\bar\rho')}{2\alpha+ 8\pi}+\frac{q q'}{4\pi r^4}=0,
\end{eqnarray}
where the gravitational mass, $M_G(r)$, is expressed by using the
following formula as
\begin{eqnarray}\label{hgg52}
M_G(r)=\frac{r B_2 A_2'}{A_2^2}.
\end{eqnarray}
Putting the above value into Eq.\eqref{hhg52} yields
\begin{eqnarray}\label{hg53}
-\frac{A_2'}{A_2}(\bar\rho+\bar p_r)-\frac{2}{r}(\bar p_r-\bar p_t)-
\bar p_r'-\frac{\alpha(\bar p_r'+2\bar p_t'+\bar\rho')}{2\alpha+
8\pi}+\frac{q q'}{4\pi r^4}=0.
\end{eqnarray}
By combining Eqs.\eqref{hg52} and \eqref{hg53}, the individual
forces are represented as
\begin{eqnarray}
&f_a&=\frac{2}{r}(\bar p_t-\bar p_r),\\ \label{hhh52} &f_h&=-\bar p_r',\\
\label{ii52} &f_g&=-\frac{A_2'}{A_2}(\bar\rho+\bar
p_r),\\\label{iii52}&f_m&=-\frac{\alpha(\bar p_r'+2\bar p_t'+\bar
\rho')}{2\alpha+8\pi}, \\\label{iii522}&f_e&=\frac{q q'}{4\pi r^4}.
\end{eqnarray}
Figure \textbf{7} depicts stability in stellar bodies by balancing
various forces. The graph shows that opposing forces result in a net
sum of zero. The star's stability and resistance to gravitational
collapse are attributed to a perfect balance of opposing forces that
maintain hydrostatic balance.
\begin{figure}[h!]\center
\epsfig{file=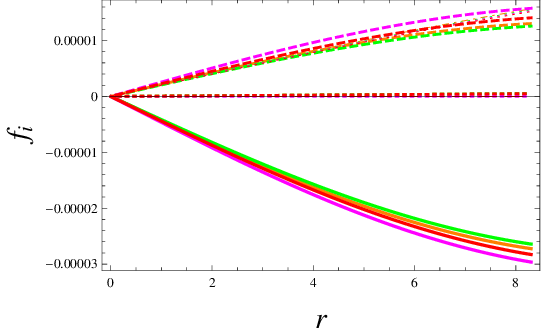,width=0.45\linewidth}\epsfig{file=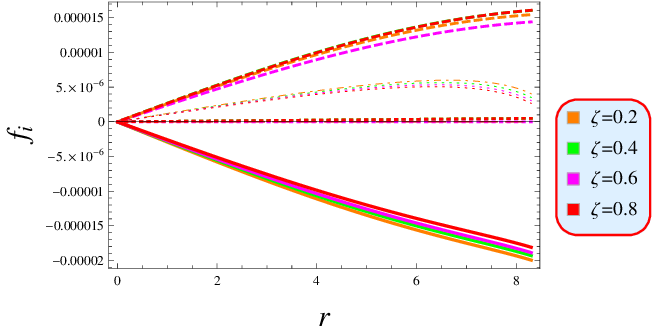,width=0.54\linewidth}
\epsfig{file=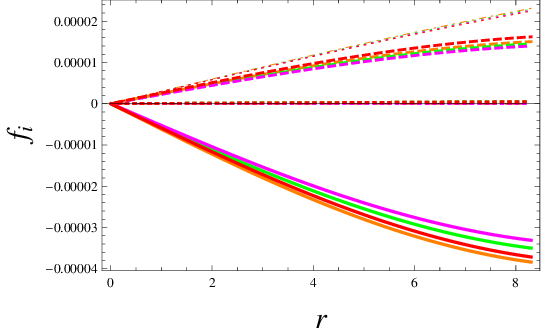,width=0.45\linewidth}\epsfig{file=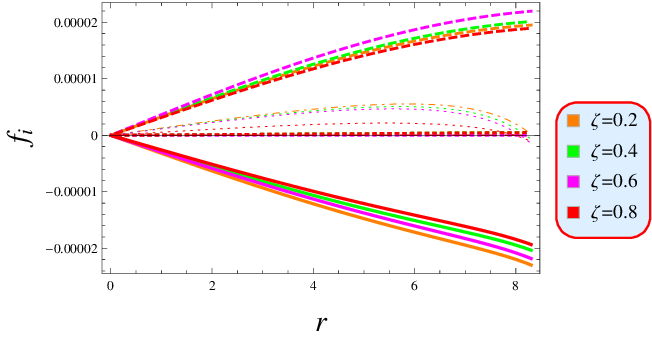,width=0.54\linewidth}
\epsfig{file=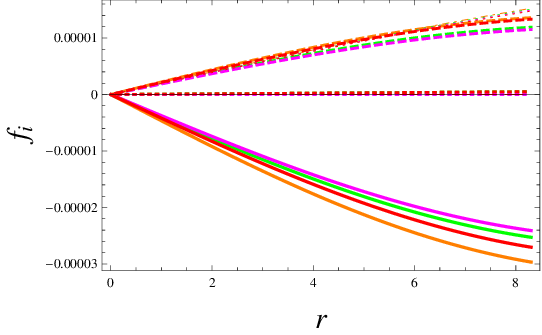,width=0.45\linewidth}\epsfig{file=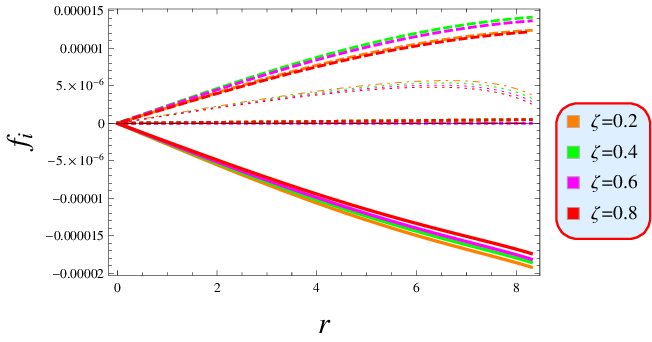,width=0.54\linewidth}
\epsfig{file=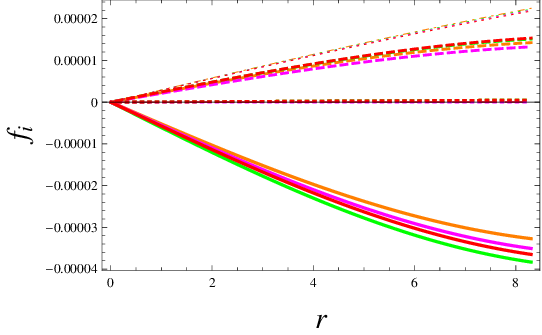,width=0.45\linewidth}\epsfig{file=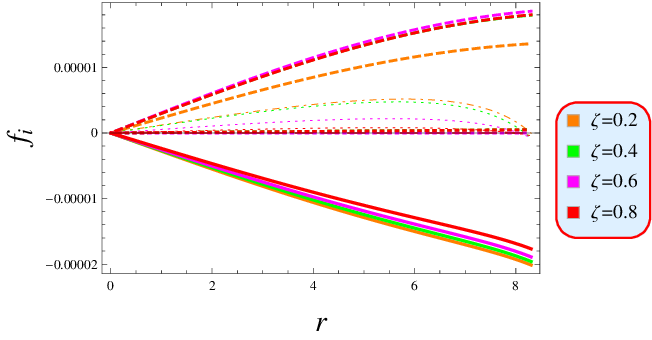,width=0.54\linewidth}
\caption{Equilibrium of forces for solutions 1 (left) and 2 (right)
for $Q=0.3$ at $\omega=0.1$ (first row) \& $0.2$ (second row), and
for $Q=0.6$ at $\omega=0.1$ (third row) \& $0.2$ (forth row). Here,
$f_g$, $f_h$, $f_a$, $f_m$ and $f_e$ correspond to thick, dashed,
dot-dashed, long dashed and dotted lines, respectively.}
\end{figure}
\begin{figure}[h!]\center
\epsfig{file=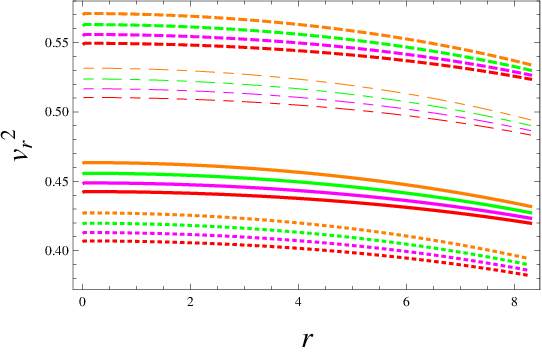,width=0.45\linewidth}\epsfig{file=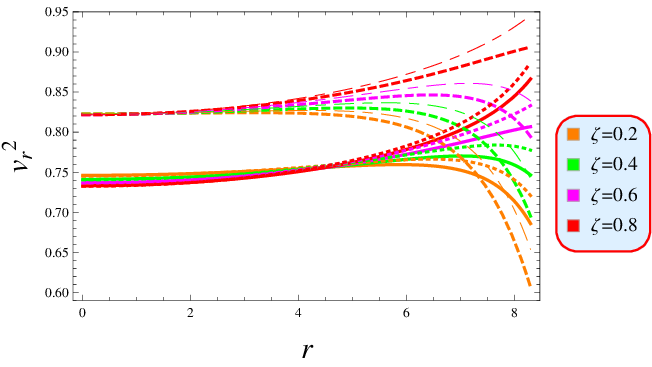,width=0.54\linewidth}
\epsfig{file=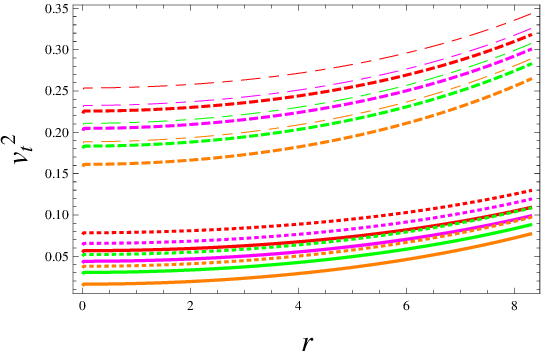,width=0.45\linewidth}\epsfig{file=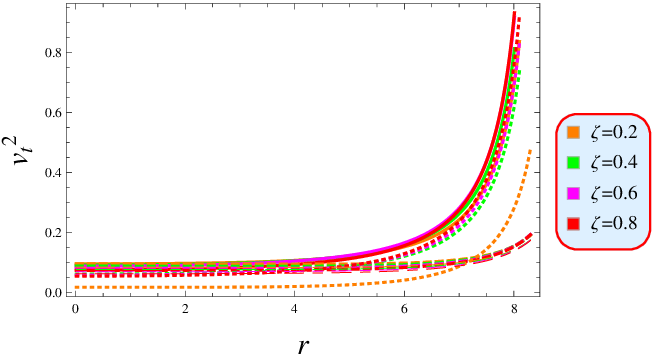,width=0.54\linewidth}
\epsfig{file=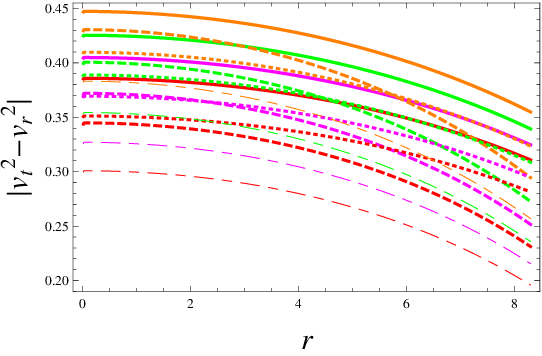,width=0.45\linewidth}\epsfig{file=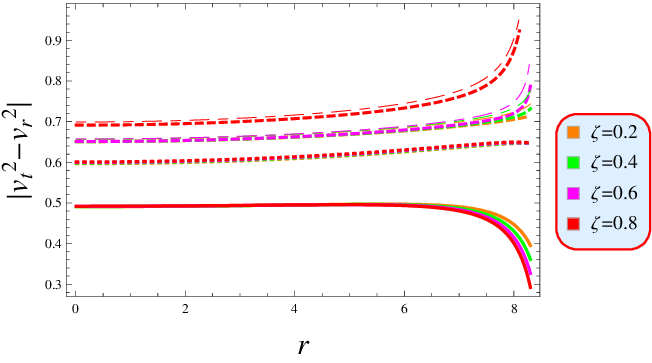,width=0.54\linewidth}
\caption{Causality and cracking for solutions 1 (left) and 2 (right)
for $Q=0.3$ at $\omega=0.1$ (thick) \& $0.2$ (dashed), and for
$Q=0.6$ at $\omega=0.1$ (dotted) \& $0.2$ (small-long dashed).}
\end{figure}

\subsubsection{Causality and Cracking}

The stability of self-gravitating stars is an important problem in
theoretical astrophysics. This study examines the radial and
non-radial speed components of acoustic waves, which are described
as $v_{r}^{2}=\frac{d\bar p_{r}}{d\bar\rho}$ and
$v_{t}^{2}=\frac{d\bar p_{t}}{d\bar\rho}$. Abreu et al. \cite{42bb}
found that a system maintains causality when the velocity of light
exceeds that of sound, meaning that $0 < v_{r}^{2}, ~ v_{t}^{2} <
1$. Figure \textbf{8} demonstrates that both produced solutions that
meet the criterion for each parameter, proving physical stability.

In materials with an unequal distribution of substance, cracking
patterns can show equilibrium. On both sides of the cracking point,
fluid elements experience different accelerations, illustrating how
the system changes as it deviates from equilibrium. Cracking in
self-gravitating compact objects was hypothesized by Herrera
\cite{42ba}. The absolute difference between the tangential and
radial sound velocities, $|v_t^2 - v_r^2|$, must fall inside the
interval [0,1] for a configuration to be stable. Both models meet
the stability condition, as shown in Figure \textbf{8}.
\begin{figure}[h!]\center
\epsfig{file=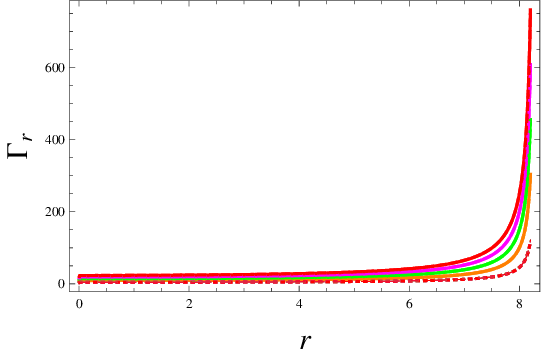,width=0.45\linewidth}\epsfig{file=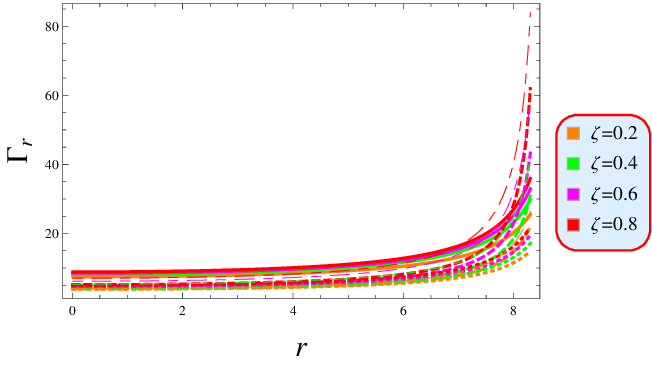,width=0.54\linewidth}
\epsfig{file=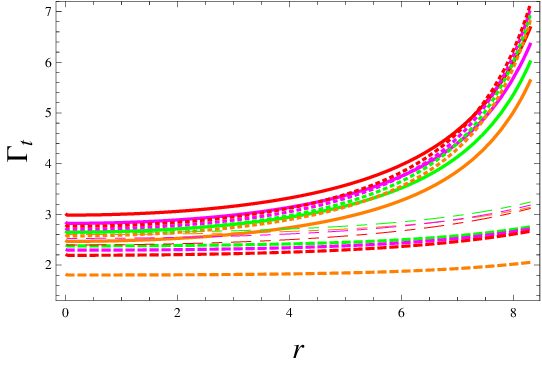,width=0.45\linewidth}\epsfig{file=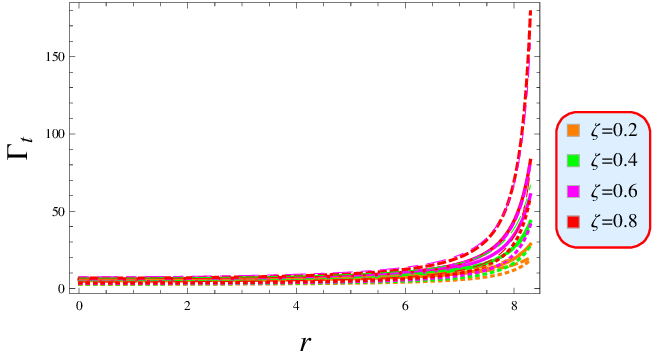,width=0.54\linewidth}
\caption{Adiabatic index for models 1 (left) and 2 (right) for
$Q=0.3$ at $\omega=0.1$ (thick) \& $0.2$ (dashed), and for $Q=0.6$
at $\omega=0.1$ (dotted) \& $0.2$ (small-long dashed).}
\end{figure}

\subsubsection{Adiabatic Index}

The structural stability of compact stars is also determined by the
adiabatic index. This dimensionless quantity helps to measure the
specific heat ratio at constant pressure and volume. It serves as a
measure for massive stars' internal particle interactions and
thermal characteristics. This index can be used by researchers to
understand more about the general structural stability of dense
stellar objects, the consistency conditions, and the link between
fluid variables. The mathematical formula of adiabatic index for the
anisotropic fluid is as follow
$$\Gamma_{r}=\frac{\bar p_{r}+\bar\rho}{\bar p_{r}}\frac{d\bar p_{r}}{d\bar\rho}, \quad
\Gamma_{t}=\frac{\bar p_{t}+\bar\rho}{\bar p_{t}}\frac{d\bar
p_{t}}{d\bar\rho}.$$ For a stable configuration, all of the factors
listed above must be bigger than $\frac{4}{3}$ \cite{42f}. Figure
\textbf{9} shows that both solutions meet the stability criteria.

\section{Discussion and Conclusions}

This study presents spherically symmetric models with two
independent fluid sources in $f(R,T)$ theory under the presence of
an electric charge. The MGD technique in this regard provides
insights into the structural evolution of massive astrophysical
objects. The impact of the decoupling parameter on star evolution
has also been studied. A new source was initially added to the seed
matter distribution and separated by a radial transformation. Both
systems are solved independently, resulting in a unique solution for
the entire configuration. The first fluid source required two
constraints, a specific radial metric function and anisotropy. The
second source was also solved by considering two constraints
independently, the relationship between $\Psi_{0}^{0}$ and seed
energy density, and the link between $\Psi_{1}^{1}$ and radial
pressure. Integrating these equations accurately yields the desired
decoupling function, which led to the components of the new source
$\Psi_{\alpha\eta}$, enabling the creation of specific interior
spacetime geometries. To compute the constants in these solutions,
we compare the inner geometries to the outer Reissner-Nordstr\"{o}m
metric at the hypersurface. The highlights of this study of both
proposed models are provided below.
\begin{itemize}
\item Our results present the acceptable behavior of metric functions
and matter components in star formations (Figures \textbf{1} and
\textbf{2}). The absence of radial pressure on the spherical surface
is also observed. Both models exhibit positively increasing
anisotropy, indicating a repulsive force. This assures stability of
our proposed configurations.

\item The regularity conditions indicate a valid pattern.
The results for redshift, compactness, and mass correspond with the
expected outcomes (Figure \textbf{3}). A clear positive linear
association exists between the moment of inertia and the total mass
(Figure \textbf{4}).

\item The equation of state indicates that variables $\omega_r$
and $\omega_t$ are both smaller than one, confirming the validity
(Figure \textbf{5}). The models are physically acceptable and meet
all energy requirements (Figure \textbf{6}). The results suggest the
presence of a standard fluid within the stars' interior.

\item Self-gravitating systems require forces' balance to function
properly. Our models gain the hydrostatic equilibrium as the sum of
all forces equals to zero (Figure \textbf{7}). The study also
analyzes models' stability using three methods: sound speed
fluctuations, cracking, and adiabatic index. All of them show that
both solutions are stable, as the said criteria are satisfied
(Figures \textbf{8} and \textbf{9}).

\item Both models meet the stability and physical requirements for
celestial objects in the $f(R,T)$ framework. This study found that
the MGD technique effectively simulates the interiors of compact
stars with anisotropic fluid distribution.

\item Notably, in our study, we have decoupled the field equations by assuming
a non-exponential interior metric and applied the binomial theorem
up to $O(\omega)$. Interestingly, our results for MGD-based
anisotropic fluid stars in $f(R,T)$ are consistent with those
presented in \cite{42fa}-\cite{42fe}.
\end{itemize}

This research has important implications for both theoretical and
observational astrophysics. To represent the complex behavior of
high-density materials in compact stars, anisotropy must be
considered with an effect of electric charge. The proposed solutions
align with the observations from established compact stars,
providing insight into mass-radius relation and inner composition.
Future research could focus on utilizing the improved equations of
state, or taking into account the magnetic field, or rotation to
capture the structural integrity. This approach is applicable to
gravity theories and models with extra spatial dimensions, revealing
previously unknown phenomena under strong gravity. This discovery
provides the way for future research on dense stars in modified
theories and their significance in the universe.
\\\\
\textbf{Data Availability Statement:} No new data is generated in
this study.

\end{document}